\documentclass[
superscriptaddress,
 amsmath,amssymb,
 aps,12pt
]{revtex4-2}
\usepackage[dvipdfmx]{graphicx}
\usepackage{ascmac}
\usepackage{amsmath}
\usepackage{amsthm}
\usepackage{bm}
\usepackage{txfonts}
\usepackage{braket}
\usepackage[top=25truemm,bottom=25truemm,left=20truemm,right=20truemm]{geometry}
\usepackage{setspace}
\usepackage{here}
\usepackage{comment}
\usepackage{color}
\newtheorem{thm}{Theorem}
\newenvironment{thmbis}
  {\addtocounter{thm}{-1}%
   \begin{thm}}
  {\end{thm}}
\allowdisplaybreaks[2]
\usepackage[normalem]{ulem}
\newcommand{\rev}[2]{#2}

\begin{document}
\title{Limitations of Quantum Measurements and Operations of Scattering Type under the Energy Conservation Law}

\author{Ryota Katsube \footnote{I have been involved in the research for this paper since I was in graduate school of Science, Tohoku University. }} 
\affiliation{Department of Physics, Graduate School of Science, Tohoku University, Sendai 980-8578, Japan}

\author{Masanao Ozawa}
\affiliation{ Center for Mathematical Science and Artificial Intelligence, Academy of Emerging Sciences,
Chubu University, 1200 Matsumoto-cho, Kasugai 487-8501, Japan}
\affiliation{Graduate School of Informatics, Nagoya University, Chikusa-ku, Nagoya 464-8601, Japan}

\author{Masahiro Hotta}
\affiliation{Department of Physics, Graduate School of Science, Tohoku University, Sendai 980-8578, Japan}
\affiliation{Leung Center for Cosmology and Particle Astrophysics, National Taiwan University,
No.1, Sec.4, Roosevelt Road, Taipei 10617, Taiwan (R.O.C)}

\date{\today}


\begin{abstract}
It is important to improve the accuracy of quantum measurements and operations both in engineering and fundamental physics.  It is known, however, that 
the achievable accuracy of measurements and unitary operations are generally limited by conservation laws according to the Wigner--Araki--Yanase theorem (WAY theorem) and its generalizations.  Although many researches have extended the WAY theorem quantitatively, most of them, as well as the original WAY theorem, concern only additive conservation laws like the angular momentum conservation law.  In this paper, we explore the limitation incurred by the energy conservation law, which is universal but is one of the non-additive conservation laws. 
We present a lower bound for the error of a quantum measurement using a scattering process satisfying the energy conservation law.  We obtain conditions that a control system Hamiltonian must fulfill in order to implement a controlled unitary gate with zero error when a scattering process is considered. We also show the quantitative relationship between the upper bound of the gate fidelity of a controlled unitary gate and the energy fluctuation of systems when a target system and a control system are both one qubit.
\end{abstract} 
\maketitle

\section{Introduction}
It is essential to improve the accuracy of quantum measurements and quantum gate operations in quantum information processing. 
Quantum computing attracts industrial attention, since Shor \cite{Shor_factor, Shor_factor2} found an efficient quantum algorithm for prime factorization, the hardness of which is assumed in some protocols for public key cryptography.
However, it is demanding to reduce the error in gate operations and measurements below the rate given by the threshold theorem for successful error correction to realize fault-torelant quantum computation \cite{Knill,Kitaev_1997,Aharonov,Fukui,Schotte}. Quantum key distribution \cite{BB84}, which aims to realize an information-theoretically secure cryptosystem, also demands to reduce the error to perform the protocol effectively, since any error, even not caused by eavesdroppers, reduces the key rate by error reconciliation.  Other examples that require high accuracy of measurements in fundamental physics are the observation of the polarization of photons in the cosmic background radiation, which is thought to carry information about the early universe \cite{Hazumi, Planck}, and the observation of gravitational waves, which has recently been used as a tool to elucidate the origin of elements and the early universe \cite{Blackhole_gw}.

On the other hand, it is known that a conservation law limits the accuracy of measurements and unitary operations. Wigner \cite{Wigner} first showed in 1952 that the projective measurement of the spin $x$-component of a particle with spin $\frac{1}{2}$ cannot be realized under the conservation of angular momentum along the $z$-axis. Later, Araki and Yanase \cite{Araki} showed the no-go theorem which states that, in general, an additive conservation law limits the \rev{accuracy}{ accurate implementation } of  projective measurement of the quantity not commuting with the conserved quantity. This theorem is called the Wigner--Araki--Yanase theorem (WAY theorem).  

Ozawa \cite{Ozawa-91CP,Ozawa-93WA,Ozawa-02CLU} obtained quantitative generalizations of the WAY theorem by introducing a systematic method of manipulating the noise commutation relations, the commutation relations between the noise operator, previously introduced by von Neumann \cite[p.~404]{Neumann55}, and the conserved quantity, and
established the tradeoff between the measurement error and the fluctuation of the conserved quantity in the apparatus for arbitrary measurements, originally suggested by Yanase \cite{Yanase} for spin measurements.
Ozawa \cite{Ozawa-ineq} also derived a universally valid reformulation of Heisenberg's error-disturbance relation and  in \cite{Ozawa-way} showed that \rev{the}{ his } quantitative generalization \cite{Ozawa-02CLU} of the WAY theorem is a 
\rev{}{ straightforward }consequence of \rev{the conservation law and Heisenberg's error-disturbance relation in its}{ his new
error-disturbance relation \cite{Ozawa-ineq}}.  

\rev{The}{Ozawa also applied the }
noise commutation relations \rev{were also used}{} to evaluat\rev{e}{ing } the accuracy of gate implementations under conservation laws for the CNOT gate \cite{Ozawa-CNOT}
and the Hadamard gate  \cite{Ozawa-way},
showing that to be the smaller the error, it is necessary to make the greater the fluctuation of the conserved quantity in the controller, or the size of the controller.  
\rev{The}{Gea-Banacloche et al. \cite{Gea-Banacloche-05CQL,Gea-Banacloche-06MEP} compared  the } 
 above approach \rev{was compared}{} with Gea-Banacloche's \cite{Gea-Banacloche-Ban02,Gea-Banacloche-Ban02b,Gea-Banacloche-BK03,Gea-Banacloche-BK05} model dependent approach to the error in quantum logic gates caused by the quantum nature of controlling systems \rev{}{to show their consistency. }
Karasawa and his collaborators \cite{Karasawa-NOT,Karasawa-1qubit} proved the implementation error 
bound for the NOT gate and later for the arbitrary one-qubit gates \rev{}{ to cover a universal set of elementary gates. } Afterwards, Tajima et al.  \cite{Tajima-measure, Tajima-coherence} provided a tighter lower bound for the error of quantum measurement and implementations of quantum gates under conservation laws by expressing the fluctuation of the conserved quantity in terms of the quantum Fisher information. More recently, Mohammady et al. \cite{Mohammady23}
derived general quantitative relations between measurement error and disturbance under
additive conservation laws based on POVMs and quantum channels.

The WAY theorem tells us that when we want to apply operations that break the symmetry, we must compensate  \rev{}{ the fluctuations of } system's conserved quantity\rev{ fluctuations}{}, i.e., the coherence with respect to conserved quantities, as a resource. This point of view has been systematized as the quantum resource theory of asymmetry. 
The operations that are feasible in the presence of symmetry have been investigated 
\rev{}{ by Bartlett, Marvian, Ahmadi, \AA{}berg, and others }
\cite{Bartlett, Marvian-resource, Ahmadi, Marvian-state, Aberg, Marvian-broadcast, Lostaglio, Miyadera-resource, Takagi-resource}.  The WAY theorem has also been applied to the black hole information loss problem \cite{Blackhole-paradox, Page, Hayden} to investigate to which degree of completeness  the scrambled information can be recovered in the presence of symmetry 
\rev{}{ by Nakata, Tajima, and others } \cite{Nakata, Tajima-blackhole}.

Although most of the work on the WAY theorem has been done on additive conservation laws, 
\rev{that do not include interaction terms, such as momentum and angular momentum conservation laws,}{}  some researchers extended the WAY theorem to \rev{multiplicative}{ non-additive } conservation laws.
\rev{}{Kimura et al. \cite{Kimura} } extended the WAY theorem to multiplicative conservation laws.
Navascu\'es et al. \cite{Navascues} proved relations between the difference between a desired POVM measurement we want to implement and an actual measurement we are able to implement under the energy conservation and the energy distribution.  Miyadera et al. \cite{Miyadera-energy}
showed an inequality between the fidelity of information distribution and the norm of the interaction term of Hamiltonian. Tukiainen \cite{Tukiainen} proved a WAY-type no-go theorem beyond conservation laws in the assumption of the weak Yanase condition.
We are able to obtain an upper bound of  gate fidelity by the WAY theorem in general.
\rev{}{Chiribella et al. \cite{Chiribella} also showed } 
that there is a lower bound of gate fidelity under the energy conservation law using methods in the resource theory.

In this paper, we \rev{prove}{ show } that \rev{the relation}{ Ozawa'a lower bound } \cite{Ozawa-02CLU} 
\rev{between a measurement error, non-commutativity of the measured quantity and the conserved quantity, and the variance of the conserved quantity is able to}{ for the measurement error under an additive conservation law can } be extended to measurements in scattering processes under energy conservation and we have a similar lower bound of the measurement error.   We also show an upper bound of the gate fidelity of SWAP gate as an application of the theorem.  Furthermore, we prove necessary conditions for the perfect implementation of a controlled unitary gate in a scattering process under energy conservation. In addition, we also show an inequality between the gate fidelity and energy variances for two-qubits controlled unitary gate, which is the extension of Ozawa's result \cite{Ozawa-CNOT}.

One of the possible applications of our results for fundamental science is a design of gravitational wave detectors.  Recently some analysis of the dynamics of the gravitational wave as a quantum field are known \cite{Maulik-detector,Maulik-gravity,Kanno}. In the observation of a gravitational wave using a laser interferometer,  the information of the  gravitational wave is transfered to the laser interferometer via the gravitational interaction.  Then we are able to know the state of the gravitational wave by measuring the state of the laser interferometer. However the energy conservation gives the constraints for measurement, we have to design detectors to lower the measurement error.

Section 2 presents a lower bound for the measurement error of quantum measurements in scattering processes, which are important in particle physics experiments, under the requirement of energy conservation.  In section 3, we show the conditions that the free Hamiltonian of the control system must satisfy in order to obtain a vanishing  implementation error of the control unitary gates  in the case of energy conservation. We also give an inequality between the energy fluctuation and the gate fidelity of two-qubit controlled unitary gates under  energy conservation in section 3. Finally, we summarize the results in section 4.


\section{Fundumental error bound of scattering quantum measurements under energy conservation law}

In this section, we present an error limit of quantum measurements of scattering type
under the energy conservation law.

Firstly, we review Ozawa's inequality \cite{Ozawa-ineq} and the quantification of the WAY theorem using it \cite{Ozawa-02CLU,Ozawa-way} for our later discussion.  Let's consider that we want to measure an observable $A_S$ of a system $S$ indirectly using a meter observable $M_D$ of a detector system $D$, in other words, we measure $A_S$ at the initial time $t=0$  by measuring the meter observable $M_D$
at $t = \tau$. In the Heisenberg picture, we indirectly measure $A_S(0)$ by measuring $M_D(\tau)= U^{\dagger}M_D U$, where $U$ is a unitary time evolution operator from time $t=0$ to $t=\tau$ that describes the process in which $S$ and $D$ become correlated. The output probability distribution of this indirect measurement is described by a POVM as follows. Let us denote a spectral decomposition  of $M_D$ by $M_D = \sum_m m P_{m,D}$, where $ P_{m,D}$ is a projection operator associated with an eigenvalue $m$ of $M_D$. Let us suppose that $S$ and $D$ are not correlated in the initial time and the initial state of the composite system $S$+$D$ is given by $\rho_S \otimes \sigma_D$.  The probability that the measured value of $M_D$ is $m$,  represented as ${\rm Pr}(M_D = m|| \rho_S) $ , is given by ${\rm Pr}(M_D = m|| \rho_S) = {\rm Tr}_{SD}[(I_S \otimes P_{m,D})U(\rho_S \otimes \sigma_D) U^\dagger ]$. Defining a POVM element $E_m$ by $E_m = {\rm Tr}_D [(I_S \otimes \sigma_D) U^\dagger (I_S \otimes P_{m,D})U] $,  \rev{}{ the relation } $ {\rm Pr}(M_D = m|| \rho_S) = {\rm Tr}_S[\rho_S E_m]$ holds.  \par

We define the error operator ${\bf E}$ for the measurement of $A_S$ and the disturbance operator ${\bf D}$ for an observable  $B_S$  as follows:
	\begin{eqnarray}
{\bf E} &=& U^{\dagger} (I_S \otimes M_D) U - A_S \otimes I_D,  \label{eq:mesurement_error_operator} \\
{\bf D} &=& U^{\dagger} (B_S \otimes I_D) U - B_S \otimes I_D.
	\end{eqnarray}
In the indirect measurement, we indirectly measure $A_S(0) = A_S \otimes I_D $ by measuring  $M_D(\tau)$ $= U^{\dagger}(I_S \otimes M_D) U$. We defined the error operator by the difference between $A_S(0)$ and $M_D(\tau)$. Similarly, we defined the disturbance operator by the difference between $B_S$ at $t= 0$ and $B_S$ at $t=\tau$. We define the measurement error $\varepsilon(A_S)$ and the disturbance $\eta(B_S)$ by $\varepsilon(A_S)= \sqrt{{\rm Tr}[{\bf E}^2 (\rho_S \otimes \sigma_D)]}$ and $\eta(B_S) = \sqrt{{\rm Tr}[{\bf D}^2 (\rho_S \otimes \sigma_D)]}$. Then the following Ozawa's inequality 
	\begin{eqnarray}
\varepsilon(A_S) \eta(B_S) + \varepsilon(A_S) \sigma(B_S) + \sigma(A_S) \eta(B_S) \geq \frac{1}{2} \left| {\rm Tr} \left( [A_S,B_S]\rho_S\right)\right| 
	\end{eqnarray}
holds \cite{Ozawa-ineq}, where $\sigma(A_S)$ and $\sigma(B_S)$ represent standard deviations of $A_S$ and $B_S$ in the state $\rho_S$, respectively.

Next, we review the lower bound of the measurement error $\varepsilon(A_S)$ when there is an additive conserved quantity \cite{Ozawa-02CLU}.  We suppose that $L = L_S \otimes I_D + I_S \otimes L_D $ is an additive conserved observable for the measurement process. For example, the angular momentum and the momentum are additive conserved observables. When we have an additive conserved quantity $L$, the time evolution operator $U$ which characterizes the measurement process  must satisfy  the following condition:
	\begin{eqnarray}
[U,L] = 0.
	\end{eqnarray}
We further assume that the commutativity of the meter observable $M_D$ and the detector's conserved quantity $L_D$, which is called Yanase's condition, i.e.,  $[M_D, L_D]=0$.  Then it is known that 
\begin{eqnarray}
\varepsilon(A_S)^2 \geq \frac{\left| {\rm Tr}\left([A_S,L_S] \rho \right) \right|^2}{4\sigma(L_S)^2 + 4\sigma(L_D)^2} \label{eq:way-review}
	\end{eqnarray}
holds \cite{Ozawa-02CLU}. In \cite{Ozawa-way}  it is shown that Eq.~(\ref{eq:way-review}) is a consequence of Ozawa's inequality \cite{Ozawa-ineq}. This enables us to evaluate the lower bound of the measurement error quantitatively when there is an additive conserved observable.

In the following, we show that Eq.~(\ref{eq:way-review}) is able to be extended to measurements in scattering processes
under energy conservation and we have a similar lower bound of the measurement error. In the following, we consider two spin $\frac{1}{2} $ particles I and II. Each of them has one degree of freedom in position space (the orbital degree) and the spin degree of freedom. We represent by $O_{\rm I}$ and $O_{\rm II}$  the orbital degrees of freedom of I and II, respectively.  We denote by $S$ and $D$ the spin degrees of freedom of I and II, respectively. The observable $A_S$ is assumed to be an operator on the spin state space of I.  Let I and II are initially spatially separated and having not interacted before. Then, as I approaches II, they start to interact, creating a correlation between I and II. Then, as I moves away from II, after a certain time the interaction stops. We can indirectly measure the system $S$ by measuring the system $D$, after it is correlated with the system $S$.  

Let $H_{{\rm I+II}}$ be the  time-independent Hamiltonian of the composite system I + II, given by 
	\begin{equation}
H_{{\rm I+II}} = H_{\rm I} \otimes I_{\rm II} + I_{\rm I} \otimes H_{\rm II} +H_{int}, 
	\end{equation}	
where $H_{\rm I}$ is the Hamiltonian of I, $H_{\rm II}$ is the Hamiltonian of ${\rm II}$ and $H_{int}$ is the interaction term between ${\rm I}$ and ${\rm II}$. We consider particular initial states in which the interaction term is not effective  in the time period $(0,\tau)$. The time evolution operator of the scattering process $ U_{{\rm I}+{\rm II}}$ is defined by
	\begin{equation}
U_{{\rm I}+{\rm II}}=e^{-i\tau H_{{\rm I}+{\rm II}}} ,
	\end{equation}
where we employed natural units, $\hbar = 1$.

Let us suppose that the initial state of ${\rm I}$ is  $ \ket{\psi}\ket{\chi}$ and that of ${\rm II}$ is  $\ket{\phi}\ket{\xi}$, where $\ket{\psi}$ and $\ket{\phi}$ are wavefunctions of the orbital degrees of ${\rm I}$ and ${\rm II}$, respectively, and moreover  $\ket{\chi}$ and $\ket{\xi}$ are state vectors of  $S$ and $D$, respectively.  
The final state of the composite system ${\rm I}$+${\rm II}$ at $t = \tau$ is $U_{{\rm I+II}}(\ket{\psi}\ket{\chi} \otimes \ket{\phi}\ket{\xi})$. We assume that particles ${\rm I}$ and ${\rm II}$ are spatially localized and separated at $t =0$, and the interaction $H_{int}$  is a short range interaction. Formally, we assume the following condition on the initial states and $H_{\rm I+II}$:  
\begin{equation}
 H_{int} (\ket{\psi}\ket{\chi} \otimes \ket{\phi}\ket{\xi}) = H_{int} U_{{\rm I}+{\rm II}} (\ket{\psi}\ket{\chi}\otimes \ket{\phi}\ket{\xi})= 0 \label{eq:assump1}.
	\end{equation}	
Note that we consider some specific initial orbital states $\ket{\psi} $ and $\ket{\phi}$ which satisfy Eq.~(\ref{eq:assump1}). We do not  impose Eq.~(\ref{eq:assump1}) for all orbital states of ${\rm I}$ and ${\rm II}$. On the other hand,  we consider arbitrary spin states  $\ket{\chi}$ for $S$ and an arbitrary but fixed state $\ket{\xi}$ for $D$ in Eq.~(\ref{eq:assump1}).

 We consider a lower bound of measurement errors when we indirectly measure an observable $A_S$ of $S$ by measuring a meter observable $M_D$ of $D$ under the energy conservation law,
	\begin{eqnarray}
[U_{{\rm I}+{\rm II}},H_{{\rm I}+{\rm II}}] = 0 \label{eq:energy_cons}.
	\end{eqnarray}

We assume that the meter observable satisfies the Yanase condition:
	\begin{eqnarray}
[H_{\rm II}, M_D] = 0 \label{eq:yanase}.
	\end{eqnarray}
This means that we can read off the value of the meter observable with zero error under the energy conservation.
We define the error of an indirect measurement of $A_S$, $\varepsilon(A_S)$ by
	\begin{equation}
\varepsilon(A_S) = \| \{U_{{\rm I+II}}^{\dagger}  \tilde{M_D} U_{{\rm I+II}} -\tilde{A_S} \} (\ket{\psi}\ket{\chi} \otimes \ket{\phi}\ket{\xi})\| , \label{eq:measurement_error_A}
	\end{equation}
where  $\tilde{A_S} = I_{O_{\rm I}} \otimes A_S \otimes  I_{O_{\rm II}} \otimes I_{D}$ and $\tilde{M_D} = I_{O_{\rm I}} \otimes I_S \otimes  I_{O_{\rm II}} \otimes M_{D}$. 
Under the above assumptions, the following theorem holds.
\begin{thm}\label{mainthm1}
Let $\ket{\psi}$ and $\ket{\phi}$ be orbital states which satisfy Eq.~(\ref{eq:assump1})  and $\ket{\chi}$ be arbitrary spin states and  $\ket{\xi}$ be an arbitrary but fixed state, then the following inequality for $\varepsilon(A_S)$ holds under the energy conservation:
	\begin{equation}
\varepsilon(A_S)^2 \geq \frac{|\braket{[ I_{o_{\rm I}} \otimes A_S ,H_{\rm I}]}|^2}{4\sigma^2(H_{I})+4\sigma^2(H_{\rm II})} \label{eq:way_energy},
	\end{equation}
where $\sigma(H_{\rm I})$ and $\sigma(H_{\rm II})$ are standard deviations of $H_{\rm I}$ and $H_{\rm II}$ on the initial state, respectively, and
$\braket{[I_{o_{\rm I}} \otimes A_S,H_{\rm I}]}$ represents the expectation value of the commutator $[I_{o_{\rm I}} \otimes A_S,H_{\rm I}]$ on $\ket{\psi}\ket{\chi}$.
\end{thm}

\begin{proof}
We prove the lower bound  Eq.~(\ref{eq:way_energy}) from Ozawa's inequality \cite{Ozawa-ineq} by extending the argument given in \cite{Ozawa-way}.
To prove  Eq.~(\ref{eq:way_energy}), we introduce another particle ${\rm III}$
\rev{}{that is assumed to measure the system I+II from time $\tau$ to $\tau'$ without affecting the
measurement of the system I by the system II from time 0 to $\tau$}.
The particle {\rm III} has the orbital degree denoted by $ O_{\rm III}$. Let the position and the momentum of particle ${\rm III}$ be $Q_{\rm III}$ and $P_{\rm III}$, respectively. \\

For the time region $(0,\tau)$, the  time-independent Hamiltonian of the composite system I+II+III,  the Hamiltonian $H_{{\rm I+II+III}}$ is given by 
	\begin{equation}
H_{{\rm I+II+III}} = H_{\rm I} \otimes I_{\rm II} \otimes I_{\rm III} + I_{\rm I} \otimes H_{\rm II}  \otimes I_{\rm III} +  I_{\rm I} \otimes I_{\rm II}  \otimes H_{\rm III} +H_{int} \otimes I_{\rm III}, 
	\end{equation}	
where $H_{\rm I}$ is the Hamiltonian of I , $H_{\rm II}$ is the Hamiltonian of ${\rm II}$, $H_{\rm III}$ is the Hamiltonian of ${\rm III}$ and $H_{int}$ is the interaction term between ${\rm I}$ and ${\rm II}$. 

For the time region $(\tau,\tau^\prime)$,
the Hamiltonian of the composite system is given by 
	\begin{equation}
H_{{\rm I+II+III}}^\prime = H_{\rm I} \otimes I_{\rm II} \otimes I_{\rm III} + I_{\rm I} \otimes H_{\rm II}  \otimes I_{\rm III} + I_{\rm I} \otimes I_{\rm II}  \otimes H_{\rm III} +I_{\rm I} \otimes H_{int}^\prime , 
	\end{equation}	
where $H^{\prime}_{int}$ is the interaction between ${\rm II}$ and ${\rm III}$ which is given by
    \begin{equation}
H^{\prime}_{int}= k I_{O_{\rm II}}  \otimes M_D \otimes P_{\rm III},
	\end{equation}
where $k\coloneqq \frac{1}{\tau^{\prime}-\tau}$. 

The time evolution operator for the time region $(0,\tau)$ is defined by
\begin{eqnarray}
U_{{\rm I}+{\rm II}+{\rm III}}= U_{{\rm I}+{\rm II}} \otimes  U_{\rm III} \label{eq:evol_decom}, \\
U_{{\rm I}+{\rm II}}=e^{-i\tau (H_{{\rm I}}+H_{\rm II}+H_{int})}, \\
U_{\rm III} = e^{-i \tau H_{\rm III}}.
\end{eqnarray}

The time evolution operator for the time region $(\tau,\tau^\prime)$ is defined by
\begin{eqnarray}
U_{{\rm I}+{\rm II}+{\rm III}}^\prime= e^{-i(\tau^\prime -\tau) H_{\rm I+II+III}^\prime}.
\end{eqnarray}

The initial state of ${\rm I}$ is  $ \ket{\psi}\ket{\chi}$ and that of ${\rm II}$ is  $\ket{\phi}\ket{\xi}$, where $\ket{\psi}$ and $\ket{\phi}$ are wavefunctions of the orbital degrees of ${\rm I}$ and ${\rm II}$, respectively, and moreover  $\ket{\chi}$ and $\ket{\xi}$ are state vectors of  $S$ and $D$, respectively. We denote the initial state of ${\rm III}$ by $\ket{\zeta}$. 

In  the following, we consider the indirect measurement model $\mathcal{M} = (\mathcal{H}_{O_{\rm III}}, \ket{\zeta}, U_{\rm I+II+III}^\prime  U_{\rm I+II+III},Q_{\rm III})$ for  measuring the observable $\tilde{A}_S$ on the Hilbert space $\mathcal{H}:=\mathcal{H}_{O_{\rm I}} \otimes \mathcal{H}_S \otimes \mathcal{H}_{O_{\rm II}} \otimes \mathcal{H}_D, $ which is the Hilbert space associated with $O_{\rm I}+S+O_{\rm I I}+D$. 

In the above model, we measure  $\tilde{A}_S$ at $t = 0$ by measuring $Q_{\rm III}$ at $t = \tau^\prime$. We assume that the period when the interaction $H_{int}^{\prime}$ exists, $\tau^{\prime} -\tau$, is very short. In this assumption, $H_{int}^{\prime}$ is very large compared to the Hamiltonian of ${\rm I+II+III}$, so  $H_{int}^{\prime}$ can be regarded as the total Hamiltonian in the time window from  
$\tau$ to 	$\tau^{\prime}$. In this assumption,
\begin{equation}
U^{\prime}_{{\rm I+II+III}}= e^{-i(\tau^\prime -\tau)(\tilde{M_D} \otimes P_{\rm III})} \label{eq:evolute_p}.
	\end{equation}

Next, we define  the quantum root mean square error $\alpha$  of measuring $\tilde{M_D}$ at $t=\tau$ by measuring $Q_{\rm III}$ at $t=\tau^{\prime}$  and we define the quantum root mean square error $\beta$ of measuring $\tilde{A_S}$  at $t=0$ by measuring $Q_{\rm III}$ at $t=\tau^{\prime}$, i.e. the measurement error for measuring $\tilde{A}_S$ by the indirect measurment model $\mathcal{M}$ as follows: 

\begin{eqnarray}
\alpha &=& \left\| [U^{\prime \dagger}_{{\rm I+II+III}} (\tilde{I} \otimes Q_{\rm III})U^{\prime}_{{\rm I}+{\rm II}+{\rm III}}-(\tilde{M_D}\otimes I_{\rm III})]\left(U_{\rm I+II}\ket{\psi}\ket{\chi}\ket{\phi}\ket{\xi}\right) \otimes U_{\rm III} \ket{\zeta}\right\|, \\
\beta &= & \left \| [U^{\dagger}_{{\rm I}+{\rm II}+{\rm III}} U^{\prime \dagger}_{{\rm I}+{\rm II}+{\rm III}} (\tilde{I} \otimes Q_{\rm III}) U_{{\rm I}+{\rm II}+{\rm III}}^\prime U_{{\rm I}+{\rm II}+{\rm III}} - (\tilde{A}_S \otimes I_{\rm III})] \ket{\psi}\ket{\chi} \ket{\phi}\ket{\xi} \ket{\zeta}  \right \|,
\end{eqnarray}
where $\tilde{I}= I_{\rm I} \otimes I_{\rm II}$ . \par

 To evaluate $\varepsilon(A_S)$ given in Eq.~(\ref{eq:measurement_error_A}), we first consider the Ozawa inequality for the indirect measurement model $\mathcal{M}$:
\begin{equation}
\beta \eta(\tilde{H}_0) + \beta \sigma(\tilde{H}_0) + \sigma(\tilde{A_S}) \eta(\tilde{H}_0) \geq \frac{1}{2} \left| \Braket{[\tilde{A_S},\tilde{H}_0]}\right| \label{eq:ozawa-ineq},
   \end{equation}
where $\tilde{H}_0 \coloneqq H_{\rm I} \otimes I_{\rm II}  + I_{\rm I} \otimes H_{\rm II}  $ is the Hamiltonian of the composite system ${\rm I+II}$ and $\eta(\tilde{H}_0)$ is the disturbance of $\tilde{H}_0$ for the measurement model $\mathcal{M}$, which is given by
\begin{eqnarray}
\eta(\tilde{H}_0) = \left\| [U_{{\rm I}+{\rm II}+{\rm III}}^{\dagger} U_{{\rm I}+{\rm II}+{\rm III}}^{\prime \dagger} (\tilde{H}_0 \otimes I_{\rm III})U_{{\rm I}+{\rm II}+{\rm III}}^\prime U_{{\rm I}+{\rm II}+{\rm III}} - (\tilde{H}_0 \otimes I_{\rm III})] \ket{\psi} \ket{\chi} \ket{\phi}\ket{\xi} \ket{\zeta}  \right\|.
\end{eqnarray}
For an observable $O$, $\sigma(O)$ represents the standard deviation of $O$ in the initial state.

From the Baker-Campbell-Hausdorff formula, 
	\begin{equation}
U^{\prime \dagger}_{{\rm I}+{\rm II}+{\rm III}}(\tilde{I} \otimes Q_{\rm III})U^{\prime}_{{\rm I}+{\rm II}+{\rm III}} =\tilde{I} \otimes Q_{\rm III}+\tilde{M_D} \otimes I_{\rm III} \nonumber
	\end{equation}
holds. Then we can calculate $\alpha$ as 
	\begin{eqnarray}
	\alpha 
&=&  \left\| Q_{\rm III}U_{\rm III} \ket{\zeta}\right\|. \label{eq:error_1}
	\end{eqnarray}
\par 
Looking at Eq.~(\ref{eq:error_1}), we find that the error of measuring $\tilde{M}_{D}$ at $t = \tau$ by $Q_{\rm III}$ at $t = \tau^\prime$ is given by the square root of the expectation value of $Q_{\rm III}^2$ on the  state of ${\rm III}$ at $t =\tau$. Therefore, for any $\varepsilon>0$, we assume $\alpha < \varepsilon$ by choosing $\ket{\zeta}$ appropriately.  
From Eq.~(\ref{eq:evolute_p}) and the Yanase condition, Eq.~(\ref{eq:yanase}),  $[U^{\prime}_{{\rm I}+{\rm II}+{\rm III}}U_{{\rm I}+{\rm II}+{\rm III}},\tilde{M_D}\otimes I_{\rm III}]=0$ holds. We also find the following relation:
	\begin{eqnarray}
\alpha 
&=&  \left\| [U_{{\rm I}+{\rm II}+{\rm III}}^\dagger U^{ \prime \dagger}_{{\rm I}+{\rm II}+{\rm III}}(\tilde{I} \otimes Q_{\rm III} )U_{{\rm I}+{\rm II}+{\rm III}}^\prime U_{{\rm I}+{\rm II}+{\rm III}} \right. \nonumber \\
& & \left. \quad \quad -
U_{{\rm I}+{\rm II}+{\rm III}}^\dagger U^{ \prime \dagger}_{{\rm I}+{\rm II}+{\rm III}} (\tilde{M_D} \otimes I_{\rm III})U_{{\rm I}+{\rm II}+{\rm III}}^\prime U_{{\rm I}+{\rm II}+{\rm III}} ]\ket{\psi} \ket{\chi} \ket{\phi} \ket{\xi}\ket{\zeta}\right\| . \label{eq:eq1}
	\end{eqnarray}

Next, we shall derive a relation between $\alpha$, $\beta$  and $\varepsilon(A_S)$.  From Eq.~(\ref{eq:evol_decom}) and  Eq.~(\ref{eq:measurement_error_A}), we find that
	\begin{eqnarray}
\varepsilon(A_S)
&=& \left\| [U^{\dagger}_{{\rm I}+{\rm II}+{\rm III}} U_{{\rm I}+{\rm II}+{\rm III}}^{\prime \dagger} (\tilde{M_D} \otimes I_{\rm III}) U_{{\rm I}+{\rm II}+{\rm III}}^\prime U_{{\rm I}+{\rm II}+{\rm III}}- (\tilde{A_S} \otimes I_{\rm III})] \ket{\psi} \ket{\chi} \ket{\phi}\ket{\xi}\ket{\zeta} \right\| \label{eq:eq2}
	\end{eqnarray}
holds, where we used $[\tilde{M_{D}} \otimes I_{\rm III},U^{\prime}_{{\rm I}+{\rm II}+{\rm III}}]=0$. From Eq.~(\ref{eq:eq1}), Eq.~(\ref{eq:eq2}) and the triangle inequality, we obtain the following relation: 
	\begin{eqnarray}
\beta &\leq &  \alpha + \varepsilon(A_S) \label{eq:error}.
	\end{eqnarray}
Next, we evaluate the disturbance $\eta(\tilde{H}_0)$. From the Yanase condition $[H_{\rm II} ,M_D]=0$, we have $[U^{\prime}_{{\rm I}+{\rm II}+{\rm III}},\tilde{H}_0\otimes I_{\rm III}]=0$, which means that $\tilde{H}_0$ is unchanged from $\tau$ to $\tau^\prime$. Then we obtain 
	\begin{eqnarray}
\eta(\tilde{H}_0)
& = &  \left\| [U_{{\rm I}+{\rm II}+{\rm III}}^{\dagger} U_{{\rm I}+{\rm II}+{\rm III}}^{\prime \dagger} (\tilde{H}_0 \otimes I_{\rm III})U_{{\rm I}+{\rm II}+{\rm III}}^\prime U_{{\rm I}+{\rm II}+{\rm III}} - (\tilde{H}_0 \otimes I_{\rm III})] \ket{\psi} \ket{\chi} \ket{\phi}\ket{\xi} \ket{\zeta}  \right\| \nonumber \\
&=& \left\| [U_{{\rm I}+{\rm II}+{\rm III}}^{\dagger} (\tilde{H}_0 \otimes I_{\rm III})U_{{\rm I}+{\rm II}+{\rm III}} - (\tilde{H}_0 \otimes I_{\rm III})] \ket{\psi} \ket{\chi} \ket{\phi}\ket{\xi} \ket{\zeta}  \right\| \nonumber \\
&=& \left\| [U_{{\rm I}+{\rm II}}^{\dagger} \tilde{H}_0 U_{{\rm I}+{\rm II}} - \tilde{H}_0 ] \ket{\psi} \ket{\chi} \ket{\phi}\ket{\xi}  \right\| \nonumber \\
&=& 0 \label{eq:disturbance}.
	\end{eqnarray}
 In the second line, we used that $[U^{\prime}_{{\rm I}+{\rm II}+{\rm III}},\tilde{H}_0\otimes I_{\rm III}]=0$. Eq.~(\ref{eq:evol_decom}) is utilized in the third line. In the final line, the assumption Eq.~(\ref{eq:assump1}) is considered.
From Eq.~(\ref{eq:ozawa-ineq}), Eq.~(\ref{eq:error}) and Eq.~(\ref{eq:disturbance}), 
	\begin{equation}
(\alpha + \varepsilon(A_S))\sigma(\tilde{H}_0) \geq \frac{1}{2} \left | \Braket{[\tilde{A_S},\tilde{H}_0]  }\right | \nonumber
	\end{equation}
is shown. Because $\alpha$ can be arbitrarily small by choosing $\ket{\zeta}$ appropriately,
\begin{equation}
\varepsilon(A_S)\sigma(\tilde{H}_0) \geq \frac{1}{2} \left | \Braket{[\tilde{A_S},\tilde{H_0}]  }\right | \nonumber
	\end{equation}
holds. Taking $[\tilde{A}_S,\tilde{H}_0]=[I_{o_{\rm I}} \otimes A_S ,H_{\rm I}] \otimes I_{\rm II}$ and $\sigma(\tilde{H}_0)^2 = \sigma(H_{\rm I})^2 +\sigma(H_{\rm II})^2 $ into account,
we get 
\begin{equation}
\varepsilon(A_S)^2 \geq \frac{\left|\Braket{[I_{o_{\rm I}} \otimes A_S ,H_{\rm I}]} \right|^2}{4\sigma(H_{\rm I})^2 + 4\sigma(H_{\rm II})^2}, \nonumber
	\end{equation} 
and  Eq.~(\ref{eq:way_energy}) is proved. 
\end{proof}

From Eq.~(\ref{eq:way_energy}), we can find that the lower bound of $\varepsilon(A_S)$ depends on the 
non-commutativity of $H_{\rm I}$ and $I_{o_{\rm I}} \otimes  A_S$, and variances of  Hamiltonians $H_{\rm I}$ and $H_{\rm II}$ on the initial state. Therefore we can decrease the bound of the measurement error by increasing the energy variance on the initial state. 

We  give a comment on a bound of the measurement error when we assume the weak Yanase condition $ [U^\dagger_{\rm I+II} \tilde{M}_D U_{\rm I+II}, H_{\rm I+II}] =0 $  instead of  the Yanase condition $[H_{\rm II} ,M_D]=0$. Tukiainen proved that if  $ [U^\dagger_{\rm I+II} \tilde{M}_D U_{\rm I+II}, H_{\rm I+II}] =0 $  is satisfied in  an accurate measurement of $A_S$, then $[I_{o_{\rm I}} \otimes A_S, \bra{\xi} \braket{\phi|H_{\rm I+II}|\phi} \ket{\xi}] =0 $ holds \cite{Tukiainen}. We are able to quantify it in the similar inequality to Eq.~(\ref{eq:way_energy}). 
\begin{thmbis}
Let $\ket{\psi}$ and $\ket{\phi}$ be orbital states which satisfy Eq.~(\ref{eq:assump1})  and $\ket{\chi}$ be arbitrary spin states and  $\ket{\xi}$ be an arbitrary but fixed state, then the following inequality for $\varepsilon(A_S)$ holds under the energy conservation and the weak Yanase condition:
\begin{eqnarray}
\varepsilon(A_S)^2 \geq \frac{ \left| \bra{\psi} \bra{\chi} [I_{o_{\rm I}} \otimes A_S, \bra{\xi}\bra{\phi} H_{\rm I+II} \ket{\phi}\ket{\xi}] \ket{\chi}\ket{\psi} \right|^2}{4\sigma(H_{\rm I+II})^2 } .\label{eq:bound-weak-yanase}
	\end{eqnarray}
\end{thmbis}

\begin{proof}
We consider the following Ozawa inequality instead of Eq.~(\ref{eq:ozawa-ineq}): 
	\begin{eqnarray}
\beta \eta(H_{\rm I+II}) + \beta \sigma(H_{\rm I+II}) + \sigma(\tilde{A}_S) \eta(H_{\rm I+II}) \geq \frac{1}{2} \left| \Braket{[\tilde{A}_S,H_{\rm I+II}]}\right| = \frac{1}{2} \left| \bra{\psi} \bra{\chi} [I_{O_{\rm I}} \otimes A_S, \bra{\xi} \bra{\phi} H_{\rm I+II} \ket{\phi}\ket{\xi}] \ket{\chi}\ket{\psi} \right| \label{eq:ozawa-ineq2}. \nonumber \\
	\end{eqnarray}
When the weak Yanase condition is satisfied, we are able to show that $ \eta(H_{\rm I+II}) =0 $ as follows:
	\begin{eqnarray}
\eta (H_{\rm I+II}) &=& \left \| [ U_{{\rm I}+{\rm II}+{\rm III}}^\dagger U_{{\rm I}+{\rm II}+{\rm III}}^{\prime \dagger} H_{\rm I+II}U_{{\rm I}+{\rm II}+{\rm III}}^\prime  U_{{\rm I}+{\rm II}+{\rm III}} - H_{\rm I+II}] \ket{\psi}\ket{\chi}\ket{\phi} \ket{\xi} \ket{\zeta} \right \| \\
&=&  \left \| [ U_{\rm I+II}^\dagger  H_{\rm I+II} U_{\rm I+II} - H_{\rm I+II}] \ket{\psi}\ket{\chi} \ket{\phi}\ket{\xi}  \right \| \\
&=& 0,
	\end{eqnarray}
where in the second line, we used $[\tilde{M}_D,H_{\rm I+II}] =0 $ which is obtained from the weak Yanase condition and the energy conservation law because  $[U^\dagger_{\rm I+II} \tilde{M}_D U_{\rm I+II}, H_{\rm I+II}] =U^\dagger_{\rm I+II}[\tilde{M}_D, H_{\rm I+II}] U_{\rm I+II} =0 $ holds.  In the third line,  the energy conservation law is considered. From Eq.~(\ref{eq:error}),  Eq.~(\ref{eq:ozawa-ineq2}) and $\eta(H_{\rm I+II})= 0$ and  setting $\alpha$ to be arbitrarily small,  we find that Eq.~(\ref{eq:bound-weak-yanase}) holds.
\end{proof}
\par
\begin{figure}[htbp]
\begin{center}
\includegraphics[width=100mm]{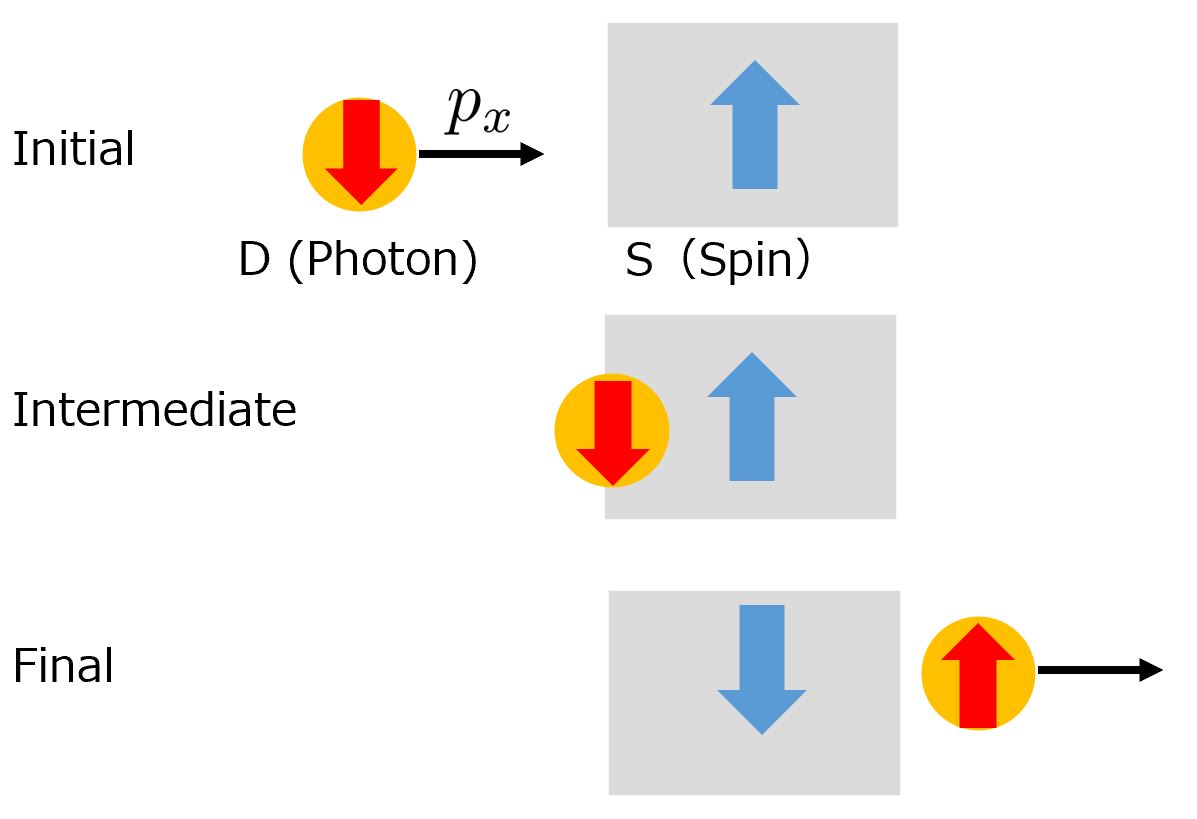}
\caption{Conceptual figure of a setting of the swapping operation. The red arrow represents the spin of the photon. The blue arrow represents the nuclear spin of the atom. The photon propagates in $x$ direction with the momentum $p_x$. After the scattering, swapping between the spins is achieved.  }
\label{fig:swap}
\end{center}
\end{figure}

Let us consider whether we are able to implement a SWAP gate using nuclear spins of heavy atoms at rest and photons as an application of Eq.~(\ref{eq:way_energy}).  We take this example because it is important from the point of view of engineering to transfer information from a  nuclear spin to a photon when we want to transmit the results of quantum computation which are encoded in states of nuclear spins by quantum communication using light.  We consider $S$ as a nuclear spin degree and  $D$ as a photon spin degree.  Let us suppose that we apply a magnetic field in the $x$ direction to the nuclear spin, and the photon moves along with the $x$-axis. There are four degrees of freedom. The first is the orbital degree of freedom of the nuclear spin. 
The second is the spin degree of freedom of the nuclear spin.  The third is the photon\rq{}s orbital degree of freedom in the $x$ direction. The forth is the spin of the photon.  We would like to swap the state of $S$ and the state of  $D$. We consider the following Hamiltonian,
\begin{eqnarray}
H_{\rm I} &=& b  \sigma_{x,S} \otimes I_D \otimes I_{o_{\rm I}} \otimes  I_{o_{\rm II}} , \\
H_{\rm II} &=&  I_S \otimes I_D \otimes I_{o_{\rm I}} \otimes  p_x , \\
H_{int} &=& g \left( \sigma_{x,S} \otimes  \sigma_{x,D} + \sigma_{y,S} \otimes \sigma_{y,D}  + \sigma_{z,S} \otimes \sigma_{z,D}  - I_S \otimes I_D \right) \otimes I_{o_{\rm I}} \otimes  w(x), \label{eq:swap_Hamiltonian}
\end{eqnarray}
where $b$ is a constant that represents the strength of the magnetic field, and $g$ is the coupling constant between the nuclear spin and the photon. In special relativity, the energy of the photon with momentum $\vec{p}$ is given by $c |\vec{p}|$, and we employed natural units, $c=1$. $w(x)$ is a function of the position operator $x$ of the photon, and it takes non-zero values only for regions where the interaction between the nuclear spin and the photon exists. This type of Hamiltonian is known as Coleman-Hepp model \cite{Hepp}. Here, we consider the dipole- dipole interaction between S and D.  This interaction has three-dimensional rotational symmetry, and the strength of the interaction $w(x)$ is able to be arbitrarily set. We consider that Eq.~(\ref{eq:swap_Hamiltonian}) is an effective Hamiltonian model which is applicable to experiments.  Fig. \ref{fig:swap} is the conceptual figure of the setting. The gray area in the figure represents the region where $w(x) \ne 0$. At first, the nuclear spin and the photon are spatially separated, and there is no interaction. When the photon approaches the nuclear spin, they begin to interact with each other.  After the scattering, the photon propagates infinity, and the interaction vanishes.  When $g$ is very large,  $H_{\rm I}$ and $H_{\rm II}$ are negligible compared to $H_{int}$ in the region where the interaction exists. In the region where $w(x)$ is able to be approximated by a constant function, we are able to make $e^{-iH_{int}t}$ be same as SWAP gate which acts on the spin state spaces of $S$ and $D$ by adjusting $gw(x)$.

We regard the implementation of SWAP gate as the measurement of a nuclear spin $\sigma_z$ where a meter observable is a photon spin $\sigma_z$ and use Eq.~(\ref{eq:way_energy}),   then we find that 
	\begin{eqnarray}
\varepsilon(\sigma_z)^2 \geq \frac{b^2 |\braket{\sigma_y}|^2}{b^2 \sigma(\sigma_x)^2 + \sigma(p_x)^2} \label{eq:swap_way}
	\end{eqnarray}
holds. Note that because $\sigma_{z,D}$ and $p_x$ are operators acting on different Hilbert spaces, the Yanase condition is satisfied. On the other hand, when we succeed in implementing SWAP gate $U_{SWAP}$ perfectly,  
	\begin{eqnarray}
\varepsilon(\sigma_z)^2 = \braket{[\sigma_{z,S} \otimes I_D - U_{SWAP}^{\dagger} (I_S \otimes \sigma_{z,D})U_{SWAP}]^2} = 0
	\end{eqnarray}
must be satisfied. Therefore, we are not able to implement SWAP gate perfectly when the fluctuation of momentum of the photon in the initial state is finite.  \par

We are able to quantify it by deriving an inequality between the gate fidelity of SWAP gate $F_{SWAP}$ and the energy variance.  Before showing the inequality of $F_{SWAP}$,  we give the definition of the gate fidelity and the worst error probability of the physical realization. We consider that we would like to implement a unitary gate $U_{ideal}$ which acts on two systems $S$ and $D$.  We represent the initial state of the composite system $S$+$D$ by  $\rho$.  We also consider an external system $E$ and suppose that its initial state is $\ket{\xi}$. Considering the purification, we are able to assume that the initial state of the external system is a pure state.  We define a unitary time evolution operator which acts on the composite system of $S$+$D$+$E$ by $U$.   The state of $S$+$D$ after $U$ is applied,  which we represent by  $\mathcal{E}(\rho)$, is  
\begin{eqnarray}
\mathcal{E}(\rho) = {\rm Tr}_E [U(\rho \otimes \ket{\xi}\bra{\xi})U^{\dagger}]. \label{eq:real_time_evolve}
\end{eqnarray}
On the other hand, if we succeed in implementing $U_{ideal} $ perfectly, the final state of $S$+$D$ is $U_{ideal} \rho U_{ideal}^{\dagger}$. Therefore, we define the worst error probability of the physical realization of $U_{ideal}$ by the completely
bounded distance (CB distance) \cite{Paulsen, Belavkin}  as follows: 
\begin{eqnarray}
D_{CB} (\mathcal{E},U_{ideal}) = \sup_{\rho_{S+D+E^{\prime}} }D\left(\mathcal{E}_{S+D}\otimes I_{E^\prime}(\rho_{S+D+E^{\prime}} ), (U_{ideal} \otimes I_{E^{\prime}}) \rho_{S+D+E^{\prime}}(U_{ideal}^{\dagger} \otimes I_{E^{\prime}})\right),
\end{eqnarray}

 where  $E^\prime$ is another external system and $D(\rho_1,\rho_2)$ is the trace distance given by
\begin{eqnarray}
D(\rho_1,\rho_2) = \frac{1}{2} {\rm Tr} \left[|\rho_1 - \rho_2| \right].
	\end{eqnarray}
The gate trace distance between $U_{ideal}$ and the physical implementation $\mathcal{E}$, $D(\mathcal{E},U_{ideal})$ is defined by
	\begin{eqnarray}
D(\mathcal{E},U_{ideal}) = \sup_{\rho_{S+D}} D(\mathcal{E}(\rho_{S+D}), U_{ideal} \, \rho_{S+D} U^{\dagger}_{ideal}).
	\end{eqnarray}
 
From the definition of the CB distance, 
\begin{eqnarray}
D_{CB} (\mathcal{E},U_{ideal}) \geq  D(\mathcal{E},U_{ideal})
	\end{eqnarray}
holds. By minimizing over all physical implementation $(U,\ket{\xi})$, 
\begin{eqnarray}
D_{CB} (\mathcal{E},U_{ideal}) \geq \inf_{(U,\ket{\xi})} D(\mathcal{E},U_{ideal})
	\end{eqnarray}
can be found. Because of the joint convexity of the trace distance, we are able to assume that $ \rho_{S+D}$ is a pure state.
The gate fidelity between $U_{ideal}$ and $\mathcal{E}$, which is denoted by $F(\mathcal{E},U_{ideal})$, is give by 
\begin{eqnarray}
F(\mathcal{E},U_{ideal}) \coloneqq \inf_{\ket{\psi}} F(\ket{\psi}),
\end{eqnarray}
where $ \ket{\psi}$ is a spin wavefunction of $S+D$ and 
\begin{eqnarray}
F(\ket{\psi}) = \braket{\psi|U_{ideal}^{\dagger} \mathcal{E}(\ket{\psi}\bra{\psi}) U_{ideal} |\psi}^{\frac{1}{2}}.
\end{eqnarray}
The following relation between the trace distance and fielity holds:
\begin{eqnarray}
D(\mathcal{E},U_{ideal}) \geq 1- F(\mathcal{E},U_{ideal})^2.
	\end{eqnarray}
Therefore, 
\begin{eqnarray}
D_{CB} (\mathcal{E},U_{ideal})  \geq D(\mathcal{E},U_{ideal}) \geq 1- F(\mathcal{E},U_{ideal})^2 
	\end{eqnarray}
is verified. If $D_{CB} (\mathcal{E},U_{ideal}) > 0$, we are not able to implement $U_{ideal}$ perfectly. \par

For the gate fidelity of SWAP gate $F_{SWAP}$, the following inequality holds.

\begin{thm}
When the Hamiltonian is given by  Eq.~(\ref{eq:swap_Hamiltonian}) , for position states which satisfy Eq.(\ref{eq:assump1}),
	\begin{eqnarray}
F_{SWAP}^2 \leq 1 - \frac{b^2}{4b^2 + 4 \sigma(p_x)^2} \label{eq:swap_fidelity}
	\end{eqnarray}
holds. 
\end{thm}

\begin{proof}
We first give a relation between  the measurement error $\varepsilon(\sigma_z)$ and $F_{SWAP}$. Then we prove an upper bound of $F_{SWAP}$ using it and Eq.~(\ref{eq:swap_way}). \par

Firstly, we write the act of $U$ in Eq.~(\ref{eq:real_time_evolve}) as
\begin{eqnarray}
U\ket{a}_S \ket{b}_{D} \ket{\xi}_E = \sum_{j,k=0}^{1}\ket{j}_S \ket{k}_{D} \ket{E_{j,k}^{a,b}}_E, 
\end{eqnarray}
where $a$ and $b$ take 0 or 1, and $\{\ket{0}, \ket{1}\}$ represents the computational basis states. From the orthonormality of the initial state,  
\begin{eqnarray}
\delta_{ac} \delta_{bd} &=& \braket{a,b|c,d} = \braket{a,b,\xi|U^{\dagger}U|c,d,\xi}  = \sum_{j,k=0}^1 \braket{E_{j,k}^{a,b}|E_{j,k}^{c,d}} \label{eq:normalortho_cond}
\end{eqnarray}
holds.  When the initial state of $S$+$D$ is $\ket{\alpha,\beta} \, (\alpha,\beta=0,1)$, the state after the time evolution described by $U$ is given by 
\begin{eqnarray}
\mathcal{E}(\ket{\alpha,\beta}\bra{\alpha,\beta}) &=& \sum_{j,k,l,m} \ket{j,k}\bra{l,m} \braket{E_{l,m}^{\alpha,\beta}|E_{j,k}^{\alpha,\beta}} .
\end{eqnarray}

Therefore, the squared of $F(\ket{\alpha,\beta})$  for SWAP gate is written as follows:
\begin{eqnarray}
F(\ket{\alpha,\beta})^2 
&=&  \braket{E_{\beta,\alpha}^{\alpha,\beta}|E_{\beta,\alpha}^{\alpha,\beta}} \label{eq:swap_pure_fidelity} .
\end{eqnarray}
\par

Next, let us  represent  the measurement error $\varepsilon(\sigma_z)$ using $\ket{E^{a,b}_{c,d}}$. We suppose that  the initial state of $S$+$D$ is $\ket{\psi} = \sum_{\alpha,\beta=0}^1c_{\alpha,\beta} \ket{\alpha,\beta}$.
We find that 
\begin{eqnarray}
\varepsilon(\sigma_z)^2 
&=& \| \left[(I_S\otimes \sigma_{z,D} \otimes I_E )U - U (\sigma_{z,S} \otimes I_{D} \otimes I_E) \right]  \ket{\psi}_{S+D} \ket{\xi}_E\|^2 \\
&=& 4 \left \| \sum_\beta c_{1,\beta} \ket{E_{0,0}^{1,\beta}}  \right\|^2 + 4 \left \| \sum_\beta c_{1,\beta} \ket{E_{1,0}^{1,\beta}}  \right\|^2 + 4 \left \| \sum_\beta c_{0,\beta} \ket{E_{0,1}^{0,\beta}}  \right\|^2  + 4 \left \| \sum_\beta c_{0,\beta} \ket{E_{1,1}^{0,\beta}}  \right\|^2  \label{eq:swap_2}
\end{eqnarray}
holds. We rewrite it by the gate fidelity. In the following, we represent  $F(\mathcal{E}, U_{SWAP})$ by $F_{SWAP}$. The following relations hold:
\begin{eqnarray}
   \varepsilon(\sigma_z)^2 
&=& 4 \left \{ |c_{1,0}|^2 \left( 1- \left\| \ket{E_{0,1}^{1,0}}  \right\|^2 - \left\| \ket{E_{1,1}^{1,0}}  \right\|^2 \right)  + |c_{0,0}|^2 \left( 1-\left\| \ket{E_{0,0}^{0,0}}  \right\|^2  - \left\| \ket{E_{1,0}^{0,0}}  \right\|^2 \right) \right. \nonumber \\
& & +2 {\rm Re} \left[ c_{1,0} c_{1,1}^* \left( \braket{E_{0,0}^{1,1}|E_{0,0}^{1,0}} +  \braket{E_{1,0}^{1,1}|E_{1,0}^{1,0}} \right)+  c_{0,0} c_{0,1}^* \left( \braket{E_{0,1}^{0,1}|E_{0,1}^{0,0}} + \braket{E_{1,1}^{0,1}|E_{1,1}^{0,0}}\right)\right] \nonumber \\
& & + \left. |c_{1,1}|^2  \left( 1 - \left \|  \ket{E_{0,1}^{1,1}}\right\|^2 - \left \|  \ket{E_{1,1}^{1,1}}\right\|^2 \right) + |c_{0,1}|^2\left( 1 - \left \|  \ket{E_{0,0}^{0,1}}\right\|^2 - \left \|  \ket{E_{1,0}^{0,1}}\right\|^2 \right)\right \} \\
& \leq & 4 \left\{ |c_{1,0}|^2 \left(1- F(\ket{1,0})^2 \right) + |c_{0,0}|^2 \left(1 - F(\ket{0,0})^2\right)\right. \nonumber \\
& & +2 \left |   c_{1,0} c_{1,1}^* \left( \braket{E_{0,0}^{1,1}|E_{0,0}^{1,0}} +  \braket{E_{1,0}^{1,1}|E_{1,0}^{1,0}}  \right)+ c_{0,0} c_{0,1}^* \left(  \braket{E_{0,1}^{0,1}|E_{0,1}^{0,0}} + \braket{E_{1,1}^{0,1}|E_{1,1}^{0,0}}\right) \right| \nonumber \\
& & + \left. |c_{1,1}|^2 \left( 1 - F(\ket{1,1})^2 \right)+ |c_{0,1}|^2 \left(1 -F(\ket{0,1})^2  \right) \right\} \\
&\leq &  4 \left\{ 1-F_{SWAP}^2+2 |c_{1,0}||c_{1,1}| \left( \left\| \ket{E_{0,0}^{1,1}} \right\|  \left\| \ket{E_{0,0}^{1,0}} \right\|+ \left\| \ket{E_{1,0}^{1,1}} \right\| \left\| \ket{E_{1,0}^{1,0}} \right\| \right)  \right. \nonumber \\
& & \left. + 2|c_{0,0}||c_{0,1}|\left( \left\| \ket{E_{0,1}^{0,1}} \right\| \left\| \ket{E_{0,1}^{0,0}} \right\|  + \left\| \ket{E_{1,1}^{0,1}} \right\| \left\| \ket{E_{1,1}^{0,0}} \right\|\right) \right\} \\
&\leq &  4 \left\{ 1-F_{SWAP}^2+ |c_{1,0}||c_{1,1}| \left( \left\| \ket{E_{0,0}^{1,1}} \right\|^2 +   \left\| \ket{E_{0,0}^{1,0}} \right\|^2+ \left\| \ket{E_{1,0}^{1,1}} \right\|^2 +\left\| \ket{E_{1,0}^{1,0}} \right\|^2 \right)  \right. \nonumber \\
& & \left. + |c_{0,0}||c_{0,1}|\left( \left\| \ket{E_{0,1}^{0,1}} \right\|^2+ \left\| \ket{E_{0,1}^{0,0}} \right\|^2  + \left\| \ket{E_{1,1}^{0,1}} \right\|^2+ \left\| \ket{E_{1,1}^{0,0}} \right\|^2\right) \right\} \\
&\leq& 4(1-F_{SWAP}^2) \left[ 1 + 2 (|c_{1,0}||c_{1,1}| + |c_{0,0}| |c_{0,1}|)\right] .
	\end{eqnarray}	

In the first line, Eq.~(\ref{eq:normalortho_cond}) is used. In second line, we considered Eq.~(\ref{eq:swap_pure_fidelity}).  In the third line,  the triangle inequality and  the Cauchy-Schwarz inequality are applied. In the fourth line, the relation between the mean square and the geometric mean is taken into consideration. In the seventh line, we used Eq.  (\ref{eq:normalortho_cond}) and Eq.~(\ref{eq:swap_pure_fidelity}).

When the Hamiltonian is given by as Eq.(\ref{eq:swap_Hamiltonian}), Eq.~(\ref{eq:swap_way}) holds and
\begin{eqnarray}
	4(1-F_{SWAP}^2) \left[ 1 + 2 (|c_{1,0}||c_{1,1}| + |c_{0,0}| |c_{0,1}|)\right]  \geq \frac{b^2 |\braket{\sigma_{y,S}}|^2}{b^2 \sigma(\sigma_{x,S})^2 + \sigma(p_x)^2} \label{eq:swap_fidelity_variance}
	\end{eqnarray}
can be proved.  Let us maximize the right-hand side over states of $S$+$D$. Because the right-hand side is independent of states of $D$, we consider maximizing over the system's pure state. It is denoted by $d_0 \ket{0} + d_1 \ket{1}$. Then we find that 
\begin{eqnarray}
	\frac{b^2 |\braket{\sigma_{y,S}}|^2}{b^2 \sigma(\sigma_{x,S})^2 + \sigma(p_x)^2} &=& \frac{4b^2 |{\rm Im}(d_0 d_1^*)|^2}{b^2(1- 4 |{\rm Re}(d_0 d_1^*)|^2)+ \sigma(p_x)^2}.	
		\end{eqnarray}
Because 
\begin{eqnarray}
	 |{\rm Re}(d_0 d_1^*)|^2 +  |{\rm Im}(d_0 d_1^*)|^2 = |d_0 d_1^*|^2  \leq \left( \frac{|d_0|^2 + |d_1|^2}{2} \right)^2 = \frac{1}{4}	\end{eqnarray}
holds,  we make the change of variales  $|{\rm Re}(d_0 d_1^*)| = r \cos \theta$, $|{\rm Im}(d_0 d_1^*)| = r \sin \theta$,  where $ 0 \leq r \leq \frac{1}{2}$,$0 \leq \theta < 2\pi$. We introduce a function $G(r,\theta)$ as follows:
\begin{eqnarray}
	G(r,\theta) = \frac{b^2 |\braket{\sigma_{y,S}}|^2}{b^2 \sigma(\sigma_{x,S})^2 + \sigma(p_x)^2} &=& \frac{4b^2r^2 \sin^2 \theta}{b^2(1- 4 r^2 \cos^2 \theta)+ \sigma(p_x)^2}.	
		\end{eqnarray}
The derivative of it is 
\begin{eqnarray}
\frac{\partial G(r,\theta)}{\partial \theta} = \frac{4b^2 r^2  \sin 2 \theta \left[b^2 (1 -4r^2) + \sigma(p_x)^2 \right]}{\left[ b^2(1- 4 r^2 \cos^2 \theta)+ \sigma(p_x)^2\right]^2}.	
	\end{eqnarray}
Since $\left[b^2 (1 -4r^2) + \sigma(p_x)^2 \right] \geq 0$ holds,  $G(r,\theta)$ takes the maximum value at $\theta = \frac{\pi}{2}$ and $\frac{3\pi}{2}$ for a fixed $ r$. Because $G\left(r,\frac{\pi}{2}\right)  = G\left(r,\frac{3\pi}{2} \right) = \frac{4b^2 r^2}{b^2 + \sigma(p_x)^2}$,  the maximum value of $G(r,\theta)$ is $\frac{b^2}{b^2 + \sigma(p_x)^2}$, which is attained at $(r,\theta) = \left(\frac{1}{2}, \frac{\pi}{2} \right) = \left(\frac{1}{2}, \frac{3\pi}{2} \right)$.  The eigenstates of $\sigma_{y,S}$ correspond to this case.

On the other hand, the coefficient of the left-hand side of Eq.~(\ref{eq:swap_fidelity_variance}), $ 1 + 2 (|c_{1,0}||c_{1,1}| + |c_{0,0}| |c_{0,1}|)$, takes the minimum value 1 when $S$'s state is the eigenstate of $\sigma_{y,S}$ and $D$'s state is $\ket{0}$. Hence, Eq.~(\ref{eq:swap_fidelity}) is derived.  On the other hand, because $p_x$ is an unbounded operator, the CB distance is 0. However if the variance of $p_x$ on the initial state is finite,  from Eq.~(\ref{eq:swap_fidelity}), we find that the implementation error occurs.     
\end{proof}
Furthermore,  we are able to extend the bound for the Hadamard gate under an additive conservation law in Ref. \cite{Ozawa-way} to that under the energy conservation law.
For example,  let us consider we would like to implement the Hadamard gate using the composite system $S$+$D$ where the Hamiltonian of the composite system is 
	\begin{eqnarray}
H = s_{x,S} + L_{x,D} + H_{int},
	\end{eqnarray} 
where $s_{x,S}$ and $L_{x,D}$ are the $x$ component of the angular momentum of each system. 
When the interaction term satisfies Eq.~(\ref{eq:assump1}) ,  from Eq.~(\ref{eq:way_energy}) and similar calculations as in Ref. \cite{Ozawa-way},  we are able to show the following bound:
\begin{eqnarray}
1 - F_H^2 \geq \varepsilon(s_{z,S})^2 \geq \frac{1}{4+ 16 \sigma(L_{x,D})^2}, \label{eq:H_fidelity}
\end{eqnarray}
where $F_H$  is the gate fidelity of the Hadamard gate.

In this section, we show the relation Eq.  (\ref{eq:way_energy}) between the measurement error in scattering processes and the energy conservation.  We also prove that SWAP gate is not able to be implemented perfectly as an application of Eq.~(\ref{eq:way_energy}). 
Furthermore,  the bound of the gate fidelity of the Hadamard gate proved in Ref. \cite{Ozawa-way} under an additive conservation law is able to be expanded under the energy conservation in scattering processes.


\section{Conditions which Hamiltonian of a controlled system must satisfy for implementing controlled unitary gates perfectly under energy conservation }

In this section, we show necessary conditions that the Hamiltonian of a control system must satisfy for implementing controlled unitary gates with zero error in scattering processes under energy conservation. We also give an upper bound of the gate fidelity of two-qubit controlled gates. 

Let us represent a control system and a target system as $C$ and $T$, respectively. We consider two orthogonal states of $C$, $\ket{\phi_0}_C$ and $\ket{\phi_1}_C$. Note that $\{ \ket{\phi_0}_C , \ket{\phi_1}_C\}$ is not necessarily a complete set. We deal with a controlled gate $U_{ideal}$ such that when the state of $C$ is $\ket{\phi_0}_C$, there is no effect on $T$, and when it is $\ket{\phi_1}_C$, a unitary operator $V_T$ is acting in $T$. The action of $U_{ideal}$ is written as 
\begin{eqnarray}
U_{ideal} \ket{\phi_0}_C \ket{\psi}_T &=& \ket{\phi_0^{\prime}}_C \ket{\psi}_T, \nonumber \\
U_{ideal} \ket{\phi_1}_C \ket{\psi}_T &=& \ket{\phi_1^{\prime}}_C \left(V_T \ket{\psi}_T \right), \label{eq:controlled-U}
	\end{eqnarray}
for an arbitrary state $\ket{\psi}_T$ of T. Note that final states of $C$, $\ket{\phi_0^{\prime}}_C$ and $\ket{\phi_1^{\prime}}_C$ are independent of the initial state of $T$, $\ket{\psi}_T$. We consider a non-trivial gate such that $V_T \ne I_T$.

Let  $H_C$, $H_T$, and $H_{int}$ denote the Hamiltonian of $C$, that of $T$, and an interaction Hamiltonian between $C$ and $T$. The total Hamiltonian of the composite system $C$ and $T$, represented as $H_{C+T}$, is given by

\begin{eqnarray}
H_{C+T} = H_C \otimes  I_T +I_C \otimes  H_T + H_{int}.
\end{eqnarray}

Similar to the previous section, we examine a scattering process in which C and T are initially spatially separated, and the interaction occurs as C approaches T. After some time, C and T become spatially separated once again. We define $\tau$ as the time interval during which the interaction is non-zero. The time evolution operator describing the scattering process $U_{CT}$ is written as
	\begin{equation}
U_{CT} = \exp(-i\tau H_{C+T}). \notag
	\end{equation}
 We assume that 
\begin{equation}
\bra{\phi}_C \bra{\psi}_TH_{int} \ket{\phi}_C \ket{\psi}_T = \bra{\phi}_C \bra{\psi}_TU_{CT}^\dagger H_{int} U_{CT} \ket{\phi}_C \ket{\psi}_T=0 \label{eq:assumption1},
	\end{equation}
for an arbitrary superposition state of $C$, $\ket{\phi}_C=c_0 \ket{\phi_0}_C + c_1 \ket{\phi_1}_C$ where $|c_0|^2 + |c_1|^2= 1$, and  an arbitrary state of $T$, $\ket{\psi}_T$.

When implementing the controlled unitary gate, $U_{ideal}$ using this scattering process, we adjust the interaction $H_{int}$ and the interaction time $\tau$ to satisfy the following relations:

\begin{eqnarray}
U_{CT} \ket{\phi_0}_C \ket{\psi}_T &=&  \ket{\phi^{\prime}_0}_C  \ket{\psi}_T, \nonumber \\
U_{CT} \ket{\phi_1}_C \ket{\psi}_T &=& \ket{\phi^{\prime}_1}_C  (V_T \ket{\psi}_T), \label{eq:ideal_ope}
	\end{eqnarray}
for arbitrary  initial state $\ket{\psi}_T$ of $T$.  However, for the Hamiltonian of C, the energy conservation gives the following constraint.

\begin{thm}

When the  energy conservation law
\begin{eqnarray}
[H_{C+T},U_{CT}] =0,
	\end{eqnarray}
holds,  $H_C$ must  satisfy the following initial condition,
\begin{equation}
\braket{\phi_0|H_C|\phi_1}_C = 0\label{eq:wayone_constraint}
	\end{equation}
to implement the gate $U_{ideal}$ with zero-error.

\end{thm}

\begin{proof}
Firstly, note that final states of $C$,  $\ket{\phi_0^{\prime}}_C$ and $\ket{\phi_1^{\prime}}_C$, are independent of the initial state of $T$,  $\ket{\psi}_T$ in Eq.~(\ref{eq:controlled-U}), and  two final states of $C$ are orthogonal to each other when we succeed in  implementing $U_{ideal}$ without error. The calculation to prove it is the same as in \cite{no-programming} in which Nielsen and Chuang proved the no-programming theorem. 
	Next  we prove that
	\begin{equation}
\bra{\phi_0}_C \bra{\psi}_T H_{int} \ket{\phi_1}_C\ket{\psi}_T =0 \label{eq:off_diagonal2}.
	\end{equation}
is valid under the assumption Eq.  (\ref{eq:assumption1}).
Substituting $\ket{\phi}_C = c_0 \ket{\phi_0}_C + c_1 \ket{\phi_1}_C$ into Eq.~(\ref{eq:assumption1}),  
 \begin{eqnarray}
 2{\rm Re}\left[c_0^* c_1 \bra{\phi_0}_C \bra{\psi}_T H_{int} \ket{\phi_1}_C\ket{\psi}_T \right] =0 \label{eq:off_diagonal1}
	\end{eqnarray} 
is obtained. Since Eq.~(\ref{eq:off_diagonal1}) holds for arbitrary complex numbers $c_0$ and $c_1$ such that $|c_0|^2 + |c_1|^2 =1$, we get Eq.~(\ref{eq:off_diagonal2}).\par

Similarly,  from the assumption of the final state in Eq.~(\ref{eq:assumption1})

\begin{eqnarray}
0 &=& \bra{\phi_0}_C \bra{\psi}_T U_{CT}^{\dagger} H_{int} U_{CT} \ket{\phi_1}_C \ket{\psi}_T \label{eq:time_evolution1}
	\end{eqnarray}

can be  shown.

Based on the above facts,  we derive Eq.~(\ref{eq:wayone_constraint}). From the orthogonality of $\ket{\phi_0}$ and $\ket{\phi_1}$,
\begin{eqnarray}
\braket{\phi_0|H_C|\phi_1}_C = \bra{\phi_0}_C \bra{\psi}_T (H_T + H_C) \ket{\phi_1}_C  \ket{\psi}_T \label{eq:wayone1}
	\end{eqnarray}
is obtained. By combining Eq.~(\ref{eq:wayone1}) with Eq.~(\ref{eq:off_diagonal2}), we get
\begin{equation}
\braket{\phi_0|H_C|\phi_1}_C = \bra{\phi_0}_C \bra{\psi}_T (H_T + H_C+H_{int}) \ket{\phi_1}_C  \ket{\psi}_T. \label{eq:wayone2}
	\end{equation}
Using Eq.~(\ref{eq:wayone2}) and the energy conservation law 

\begin{eqnarray}
[H_{C+T},U_{CT}] =0,
	\end{eqnarray}

we find that 

\begin{eqnarray}
\braket{\phi_0|H_C|\phi_1}_C &=& \bra{\phi_0}_C \bra{\psi}_T U_{CT}^{\dagger}(H_T + H_C+H_{int})U_{CT} \ket{\phi_1}_C  \ket{\psi}_T \nonumber \\
&=& \bra{\phi_0^{\prime}}_C \bra{\psi}_T (H_T + H_C +H_{int}) \ket{\phi_1^{\prime}}_C (V_T \ket{\psi}_T) \label{eq:wayone3}
    \end{eqnarray}

holds. In addition to this, from $\braket{\phi_0^{\prime}|\phi_1^{\prime}}=0$ and  Eq.~(\ref{eq:time_evolution1}),  
\begin{equation}
\braket{\phi_0|H_C|\phi_1}_C=\braket{\phi^{\prime}_0|H_C|\phi^{\prime}_1}_C \braket{\psi|V_T|\psi}_T \label{eq:wayone4}
	\end{equation}
is able to be proved for any $\ket{\psi}_T$. 

If $\braket{\phi_0|H_C|\phi_1}_C\ne 0$ holds,  $\braket{\phi^{\prime}_0|H_C|\phi^{\prime}_1}_C\ne 0$ is obtained because $V_T\ne 0$. Therefore, from Eq.~(\ref{eq:wayone4}),  we can calculate as follows:
\begin{equation}
 \braket{\psi|V_T|\psi}_T=\frac{\braket{\phi_0|H_C|\phi_1}_C}{\braket{\phi^{\prime}_0|H_C|\phi^{\prime}_1}_C}. \label{eq:wayone5}
	\end{equation}
Let the spectrum decomposition of $V_T$ be 
\begin{eqnarray}
V_T = \sum_i \alpha_i \ket{i}\bra{i},
\end{eqnarray}
where $\alpha_i$ is the $i$-th eigenvalue and $\ket{i}$ is the corresponding eigenvector. 

By substituting $\ket{\psi} = \ket{i}$ into Eq.~(\ref{eq:wayone5}), we find that
\begin{equation}
 \braket{i|V_T|i}_T=\alpha_i =  \frac{\braket{\phi_0|H_C|\phi_1}_C}{\braket{\phi^{\prime}_0|H_C|\phi^{\prime}_1}_C} \nonumber
	\end{equation}
holds for any $i$. Hence,
 \begin{eqnarray}
V_T &= &\frac{\braket{\phi_0|H_C|\phi_1}_C}{\braket{\phi^{\prime}_0|H_C|\phi^{\prime}_1}_C}  I_T \label{eq:wayone6}
	\end{eqnarray}
is obtained. Therefore, when we want to implement a non-trivial controlled unitary gate like a CNOT gate without error under energy conservation, we proved that Eq.~(\ref{eq:wayone_constraint}) has to be satisfied.

\end{proof}

Further, the following theorem holds.
\begin{thm}
When  $H_{int}$ satisfies the conditions below  similar to conditions Eq.~(\ref{eq:assump1}):
\begin{eqnarray}
H_{int} \ket{\phi_0}_C \ket{\psi}_T &=& 0,  \nonumber \\
H_{int} \ket{\phi_1}_C \ket{\psi}_T &=& 0, \nonumber \\
H_{int} \ket{\phi^{\prime}_0}_C \ket{\psi}_T &=& 0,  \nonumber \\
H_{int} \ket{\phi^{\prime}_1}_C \ket{\psi}_T &=& 0 ,  \label{eq:assump2}
	\end{eqnarray}
for an arbitrary state $\ket{\psi}_T$, the additional condition,
\begin{eqnarray}
\braket{\phi_0|H_C^2|\phi_1} =0, \label{eq:waytwo_contraint}
	\end{eqnarray}
is requested to implement $U_{ideal}$ perfectly under energy conservation. 
\end{thm}
Note that if Eq.~(\ref{eq:assump2}) holds, Eq.~(\ref{eq:assumption1}) is also satisfied. The converse is not true. Thus Eq.~(\ref{eq:assump2})  is the stronger condition than Eq.~(\ref{eq:assumption1}).  When we assume Eq.~(\ref{eq:assump2}) , we are able to show a more stringent condition Eq.(\ref{eq:waytwo_contraint}) in addition to Eq.~(\ref{eq:wayone_constraint}) .

\begin{proof}

We derive Eq.~(\ref{eq:waytwo_contraint}) under the assumption Eq.~(\ref{eq:assump2}).  A calculation similar to the derivation of Eq.~(\ref{eq:wayone4}) shows that 
\begin{eqnarray}
\braket{\phi_0|H_C^2|\phi_1}_C &=&  \bra{\phi_0}_C \bra{\psi}_T  (H_C^2 + H_T^2) \ket{\phi_1}_C \ket{\psi}_T   \nonumber \\
&=& \bra{\phi_0}_C \bra{\psi}_T [(H_C +H_T+H_{int})^2 -2 H_C \otimes H_T] \ket{\phi_1}_C \ket{\psi}_T \nonumber \\
&=& \bra{\phi_0}_C \bra{\psi}_T [U_{CT}^{\dagger}(H_C +H_T+H_{int})^2 U_{CT} -2 H_C \otimes H_T] \ket{\phi_1}_C \ket{\psi}_T  \nonumber  \\
&=& \bra{\phi^{\prime}_0}_C \bra{\psi}_T (H_C+H_T)^2 \ket{\phi^{\prime}_1}_C (V_T\ket{\psi}_T) -2\braket{\phi_0|H_C|\phi_1}_C \braket{\psi|H_T|\psi}_T \nonumber \\
&=& \braket{\phi^{\prime}_0|H_C^2|\phi^{\prime}_1}_C \braket{\psi|V_T|\psi}_T \label{eq:waytwo1}
	\end{eqnarray}
is valid for any $\ket{\psi}_T$. In the first line, we used $\braket{\phi_0|\phi_1}=0$.  We applied Eq.~(\ref{eq:assump2}) to the second  line.  In the third line,  the energy conservation is considered.  In the fourth line,  Eq.~(\ref{eq:assump2})  and Eq.~(\ref{eq:ideal_ope}) are utilized. The final line is derived by combining $\braket{\phi_0^{\prime}|\phi_1^{\prime}}=0$ and Eq.~(\ref{eq:wayone_constraint}) with Eq.~(\ref{eq:wayone4}). 

If $\braket{\phi_0|H_C^2|\phi_1} \ne 0$ works, we are able to prove that $V_T$ must be proportional to $I_T$ by a calculation similar to the derivation of  Eq.  (\ref{eq:wayone6}). Thus, we find that  if the assumption Eq.  (\ref{eq:assump2}) holds,  Eq.~(\ref{eq:waytwo_contraint}) has to be fulfilled in addition to Eq.  (\ref{eq:wayone_constraint}) to implement a non-trivial controlled unitary gate perfectly under energy conservation.    
\end{proof}
Furthermore, if $C$ and $T$ are one-qubit systems, $U_{ideal} $ becomes a two-qubit controlled gate, and $V_T$ becomes a one-qubit gate. The gate $V_T$ can be represented as 
\begin{equation}
V_T(\theta) = e^{i\phi} \left( \cos \frac{\theta}{2} I_T + i \sin \frac{\theta}{2} \vec{u} \cdot \vec{\sigma}\right), \label{eq:v_t}
	\end{equation}
where $\vec{u}$ is a three-dimensional real vector and $\vec{\sigma}$ is the vector defined by $\vec{\sigma} \coloneqq (\sigma_x, \sigma_y, \sigma_z)$, where $\sigma_x$, $\sigma_ y$ and $\sigma_z$ are the Pauli matrices which act on the Hilbert space of $T$. The domain of $\theta$ and $\phi$ are $0 \leq \phi < 2 \pi$ and $0 \leq \theta \leq \pi$, respectively. \\

For a two-qubit controlled unitary gate $U_{ideal}$, the following upper bound of the CB distance  between $U_{ideal}$ and the physical implementation $\mathcal{E}_{\alpha}$ holds.

\begin{thm}
When implementing a two-qubit controlled unitary gate  $U_{ideal}$ using a scattering process under energy conservation,  we have the following inequality regarding the CB distance between $U_{ideal}$ and the physical implementation $\mathcal{E}_{\alpha}$:
\begin{equation}
D_{CB}(\mathcal{E}_{\alpha},U_{ideal})^2 \geq \frac{\sin^2 \frac{\theta}{2}\left\|\left[\sigma(\vec{l})_C,H_C\right]\right\|^2}{16(2\gamma+\|H_A\|)^2} \label{eq:1bit}
\end{equation}
where $\sigma(\vec{l})\coloneqq \ket{\phi_0}\bra{\phi_0}-\ket{\phi_1}\bra{\phi_1}$, $\gamma \coloneqq \max\{\|H_{C}\|,\|H_T\|\}$ and $H_A$ is the free Hamiltonian of the ancillary system $A$. 

\end{thm}
If $\braket{\phi_0|H_C|\phi_1}=0$ holds, the Hamiltonian $H_C$ commutes with $\sigma(\vec{l})$ and the upper bound of gate fidelity is 1, that is to say, we can implement $U_{ideal}$ perfectly.  We also find that the bound dependence of the angle $\theta$ is $\sin^2 \frac{\theta}{2}$. \par

\begin{proof}

In the following, we denote a three-dimensional real vector which is orthogonal to $\vec{u}$, where $\vec{u}$ is given in Eq.~(\ref{eq:v_t}),  by  $\vec{v}$.  There are two vectors orthogonal to $\vec{u}$,  we set $\vec{v}$ to the vector which faces the $z$-axis positive when  we rotate $\vec{u}$ so that it faces the $x$-axis positive.   We define $\sigma(\vec{v})$ by $\sigma(\vec{v})\coloneqq \vec{v} \cdot \vec{\sigma}$. Let us denote the eigenvector with eigenvalue 1 and  the eigenvector with eigenvalue -1  by $\ket{\chi_0}$ and $\ket{\chi_1}$, respectively.
We also define $\sigma(\vec{l})$ and $\sigma(\vec{l^{\prime}})$ by $\sigma(\vec{l})\coloneqq \ket{\phi_0}\bra{\phi_0}-\ket{\phi_1}\bra{\phi_1}$ and $\sigma(\vec{l^{\prime}})\coloneqq \ket{\phi_0^{\prime}}\bra{\phi_0^{\prime}}-\ket{\phi_1^{\prime}}\bra{\phi_1^{\prime}}$. We represent the operation of $U$ as

\begin{equation}
U\ket{\phi_a}_C\ket{\chi_b}_T \ket{\xi}_A = \sum_{c,d=0}^1 \ket{\phi_c^{\prime}}_C \ket{\chi_d}_T \ket{E_{c,d}^{a,b}}_A \quad (a,b =0,1).
\end{equation}
From the orthogonality and normalization conditions,  we find that 
\begin{eqnarray}
\delta_{a,c}\delta_{b,d} = \bra{\phi_a}_C \bra{\chi_b}_T \bra{\xi}_A U^{\dagger}U\ket{\phi_c}_C \ket{\chi_d}_T \ket{\xi}_A
= \sum_{j,k} \braket{E_{j,k}^{a,b}|E_{j,k}^{c,d}} 
\end{eqnarray}
holds.
The state of $C$+$T$ after the time evolution described by $U$ is
\begin{equation}
\mathcal{E}_{\alpha}(\ket{\phi_a}_C\ket{\chi_b}_T\bra{\phi_a}_C \bra{\chi_b}_T) = \sum_{i,j,k,l} \ket{\phi_i^{\prime}}_C \ket{\chi_j}_T\braket{E_{k,l}^{a,b}|E_{i,j}^{a,b}} \bra{\phi_k^{\prime}}_C \bra{\chi_l}_T \quad (a,b =0,1).
\end{equation}
In the following,  $\ket{\phi_a}_C\ket{\chi_b}_T$ and $\ket{\phi_c^{\prime}}_C\ket{\chi_d}_T$ are  abbreviated as $\ket{a,b}$ and $\ket{c^{\prime},d}$, respectively. We distinguish the basis of $C$ with and without the prime symbol.

Since $\vec{u}$ and $\vec{v}$ are orthogonal each other,  we get
\begin{equation}
(\vec{u} \cdot \vec{\sigma})_T  \ket{\chi_0}_T = \ket{\chi_1}_T, \quad (\vec{u} \cdot \vec{\sigma})_T \ket{\chi_1}_T = \ket{\chi_0}_T.
	\end{equation}
Therefore,  
\begin{eqnarray}
U_{ideal} \ket{a,b} 
&=& \ket{0^{\prime},b} \delta_{a,0} + e^{i\phi} \left( \cos \frac{\theta}{2} \ket{1^{\prime},b} + i \sin \frac{\theta}{2} \ket{1^{\prime},1 \oplus b} \right)\delta_{a,1}
	\end{eqnarray}
can be shown.  When the initial state is $\ket{a,b}$, the gate fidelity squared is calculated as
\begin{eqnarray}
F(\ket{a,b})^2 &=& \left[  \bra{0^{\prime},b} \delta_{a,0} + e^{-i\phi} \left( \cos \frac{\theta}{2} \bra{1^{\prime},b} -i \sin \frac{\theta}{2} \bra{1^{\prime},1 \oplus b} \right)\delta_{a,1} \right]\left[ \sum_{i,j,k,l} \ket{i^\prime,j}\braket{E_{k,l}^{a,b}|E_{i,j}^{a,b}} \bra{k^{\prime},l}  \right] \nonumber \\
& & \quad  \cdot\left[ \ket{0^{\prime},b} \delta_{a,0} + e^{i\phi} \left( \cos \frac{\theta}{2} \ket{1^{\prime},b} + i \sin \frac{\theta}{2} \ket{1^{\prime},1 \oplus b} \right)\delta_{a,1} \right] .
\end{eqnarray}
When $a =0$, it is rewritten as 
\begin{eqnarray}
F(\ket{0,b})^2 = \braket{E_{0,b}^{0,b}|E_{0,b}^{0,b}},
	\end{eqnarray}
and when $ a=1$,  
\begin{eqnarray}
F(\ket{1,b})^2 
&=& \left [\cos \frac{\theta}{2}  \bra{E_{1,b}^{1,b}}  +i \sin \frac{\theta}{2} \bra{E_{1,b+1}^{1,b}} \right]  \left [\cos \frac{\theta}{2}  \ket{E_{1,b}^{1,b}}  -i \sin \frac{\theta}{2} \ket{E_{1,b+1}^{1,b}} \right] 
	\end{eqnarray}
is derived. 
For later convenience, we calculate the gate fidelity of $\tilde{U}_{ideal}= \ket{0}\bra{0}_C \otimes V_T(\pi+ \theta) + \ket{1}\bra{1}_C \otimes V_T(\pi)$. Since the gate fidelity is invariant under the unitary transformation $V_T(\pi+\theta)$ on initial states, the gate fidelity of $\tilde{U}_{ideal}$ is same as  that of $U_{ideal}$ where the gate operated  to the  target bit is $V_{T}(-\theta)$ when $C$ is on-state. When the initial state is $\ket{a,b}$, the operation of $\tilde{U}_{ideal}$ is given by 
\begin{eqnarray}
\tilde{U}_{ideal} \ket{a,b}
&=& e^{i \phi} \left( - \sin \frac{\theta}{2}  \ket{0,b} + i \cos \frac{\theta}{2}  \ket{0^{\prime},b \oplus 1} \right) \delta_{a,0} + i e^{i\phi} \ket{1^{\prime},b\oplus 1} \delta_{a,1}.
	\end{eqnarray}
 When the initial state is $\ket{a,b}$, we denote  the gate fidelity of $\tilde{U}_{ideal}$ by $f(\ket{a,b})$. 
 $f(\ket{a,b})^2$ is obtained as follows:
\begin{eqnarray}
f(\ket{a,b})^2 &=& \left[\left( - \sin \frac{\theta}{2}  \bra{0,b} - i \cos \frac{\theta}{2}  \bra{0^{\prime},b \oplus 1} \right) \delta_{a,0} - i \bra{1^{\prime},b\oplus 1} \delta_{a,1}  \right] \left[ \sum_{i,j,k,l} \ket{i^\prime,j}\braket{E_{k,l}^{a,b}|E_{i,j}^{a,b}} \bra{k^{\prime},l}  \right] \nonumber \\
& & \cdot  \left[ \left( - \sin \frac{\theta}{2}  \ket{0,b} + i \cos \frac{\theta}{2}  \ket{0^{\prime},b \oplus 1} \right) \delta_{a,0} + i  \ket{1^{\prime},b\oplus 1} \delta_{a,1} \right] .
	\end{eqnarray}
When $a =0$,  it becomes 
\begin{eqnarray}
f(\ket{0,b})^2
&=&  \left[ \sin \frac{\theta}{2} \bra{E_{0,b}^{0,b}} - i \cos \frac{\theta}{2} \bra{E_{0,b+1}^{0,b}}  \right] \left[ \sin \frac{\theta}{2} \ket{E_{0,b}^{0,b}} + i \cos \frac{\theta}{2} \ket {E_{0,b+1}^{0,b}}  \right] ,
	\end{eqnarray}
and when $a = 1$, we get
\begin{eqnarray}
f(\ket{1,b})^2 &=&\Braket{ E_{1,b+1}^{1,b}|E_{1,b+1}^{1,b}}.
	\end{eqnarray}

Next, we define the error operator $D_{CC}$ and $D_{CT}$ by
\begin{eqnarray}
D_{CC} &\coloneqq& \sin \frac{\theta}{2} \sigma(\vec{l^{\prime}})_C - \sin \frac{\theta}{2}\sigma(\vec{l})_C, \\
D_{TC} &\coloneqq &\left[ \sin \frac{\theta}{2} \sigma(\vec{v})_T - \cos \frac{\theta}{2} \sigma(\vec{v} \times \vec{u})_T \right]^{\prime} -\sin \frac{\theta}{2} \sigma(\vec{l})_C ,
	\end{eqnarray}
where for an operator $O$, we defined  $O^{\prime} \coloneqq U^{\dagger} O U$. We also denote  mean errors on $\ket{\psi}_C\ket{0}_T\ket{\xi}_A$ by $\delta_{iC} \, (i = C,T)$. They are given by
\begin{equation}
\delta_{iC}(\ket{\psi}) = \braket{D_{iC}^2}^{\frac{1}{2}} \quad (i=C,T).
\end{equation}
We first look into properties of error operators when we succeed in implementing $U_{ideal}$ perfectly, that is to say, $U = U_{ideal} \otimes I_A$.
Since 
\begin{eqnarray}
U_{ideal}^{\dagger} \sigma(\vec{l^{\prime}})_C U_{ideal} = \sigma(\vec{l})_C \otimes I_T
\end{eqnarray}
holds, $\delta_{CC}^2(\ket{\psi})=0$ is obtained  for an arbitrary $\ket{\psi}_C$ when the ideal time evolution is realized. We also find the following equation:
\begin{eqnarray}
U_{ideal}^{\dagger} \left[ \sin \frac{\theta}{2} \sigma(\vec{v})_T - \cos \frac{\theta}{2} \sigma (\vec{v} \times \vec{u})_T\right] U_{ideal}&=&\ket{0}\bra{0}_C \otimes \left[ \sin \frac{\theta}{2} \sigma(\vec{v})_T - \cos \frac{\theta}{2} \sigma (\vec{v} \times \vec{u})_T\right] \nonumber \\
& & \quad
+ \ket{1}\bra{1}_C \otimes \left[ -\sin \frac{\theta}{2} \sigma(\vec{v})_T - \cos \frac{\theta}{2} \sigma (\vec{v} \times \vec{u})_T\right] . \nonumber \\
	\end{eqnarray}
When the target state is $\ket{0}_T$ and $U=U_{ideal} \otimes I_A$, the expectation value of $D_{TC}$ is calculated as 
\begin{eqnarray}
\bra{\psi}_C \bra{0}_T D_{TC} \ket{\psi}_C \ket{0}_T 
=0,
	\end{eqnarray}
for any $\ket{\psi}_C$. Thus, when we implement $U_{ideal}$ without error,  the expectation value of error operators on $\ket{\psi}_C \ket{0}_T$ are 0. Calculating the mean square of $D_{TC}$ on $ \ket{\psi}_C \ket{0}_T$ for the case where $U= U_{ideal} \otimes I_A$,  we obtain
\begin{eqnarray} 
& &	\bra{\psi}_C \bra{0}_T D_{TC}^2 \ket{\psi}_C \ket{0}_T = \cos^2 \frac{\theta}{2}.
	\end{eqnarray}
 This expression becomes 0 for $\theta= \pi$.\par

Next we associate $\delta_{CC}^2(\ket{\psi})$ and $\delta_{TC}^2(\ket{\psi})$ with the uncertainty relation.
The energy conservation law is represented as
\begin{equation}
[U, H_T + H_C +H_A +H_{int}] =0,
\end{equation}
where $H_T$, $H_C$ and $H_A$ are the Hamiltonian of $T$, $C$ and $A$, respectively, and $H_{int}$ is the interaction term.  From the energy conservation and the triangle inequality, 
\begin{eqnarray}
\left|\Braket{[\sin \frac{\theta}{2} \sigma(\vec{l})_C,H_C]}\right| \leq  \left|\Braket{[\sin \frac{\theta}{2} \sigma(\vec{l})_C,H_C^{\prime}]} \right|+ \left|\Braket{[\sin \frac{\theta}{2} \sigma(\vec{l})_C,H_T^{\prime}]}\right|+ \left |\Braket{[\sin \frac{\theta}{2}\sigma(\vec{l})_C,H_A^{\prime}]} \right| \label{eq:1}
\end{eqnarray}
is proved for any $\ket{\psi}_C \ket{\phi}_T$. We used the assumption of the interaction term.  From the definition of error operators, we also find that 
\begin{eqnarray}
\left|\Braket{[\sin \frac{\theta}{2}\sigma(\vec{l})_C,H_C^{\prime}]}\right| &=&\left|\Braket{[H_C^{\prime},D_{TC}]}\right| ,\nonumber \\
\left|\Braket{[\sin \frac{\theta}{2}\sigma(\vec{l})_C,H_T^{\prime}]} \right|&=&\left |\Braket{[H_T^{\prime},D_{CC}]} \right|, \nonumber \\
\left |\Braket{[\sin \frac{\theta}{2}\sigma(\vec{l})_C,H_A^{\prime}]} \right | &=&  \left |\Braket{[H_A^{\prime},D_{TC}]} \right| =  \left|\Braket{[H_A^{\prime},D_{CC}]} \right| \label{eq:2}
\end{eqnarray}
hold. From Eq.~(\ref{eq:1}) and Eq.~(\ref{eq:2}),  we can show that
\begin{equation}
\left|\Braket{[\sin \frac{\theta}{2}\sigma(\vec{l})_C,H_C]}\right| \leq \left |\Braket{[H_C^{\prime},D_{TC}]} \right| + \left|\Braket{[H_T^{\prime},D_{CC}]} \right |+ \left |\Braket{[H_A^{\prime},D_{TC}]} \right| .
\end{equation}
 Using Robertson's uncertainty relation and $\Delta(D_{iC})^2 \leq \delta_{iC}^2$, where $\Delta(A)$ is the standard deviation of $A$,  we obtain
\begin{equation}
\left|\Braket{[\sin \frac{\theta}{2}\sigma(\vec{l})_C,H_C]}\right| \leq 2\delta_{TC}(\ket{\psi}) \Delta(H_C^{\prime})+2\delta_{CC}(\ket{\psi}) \Delta(H_T^{\prime}) +2\delta_{TC}(\ket{\psi}) \Delta(H_A^{\prime}) \label{eq:3}.
\end{equation}
Similarly, 
\begin{equation}
\left|\Braket{[\sin \frac{\theta}{2}\sigma(\vec{l})_C,H_C]}\right| \leq 2\delta_{TC}(\ket{\psi}) \Delta(H_C^{\prime})+2\delta_{CC}(\ket{\psi}) \Delta(H_T^{\prime}) +2\delta_{CC}(\ket{\psi}) \Delta(H_A^{\prime}) \label{eq:4}
\end{equation}
can be proved. 
Adding Eq.~(\ref{eq:3}) to Eq.~(\ref{eq:4}),
\begin{equation}
\left|\Braket{[\sin \frac{\theta}{2}\sigma(\vec{l})_C,H_C]}\right| \leq [ \delta_{TC} (\ket{\psi}) +  \delta_{CC} (\ket{\psi})] (2\gamma+\Delta(H_A^{\prime}))
\end{equation}
is derived,  where $\gamma \coloneqq \max\left \{ \|H_C\|, \|H_T\| \right \}$. Moreover, using the relation $\frac{(x+y)^2}{2} \leq x^2 + y^2$,  we get

\begin{equation}
\frac{\left|\Braket{[\sin \frac{\theta}{2}\sigma(\vec{l})_C,H_C]}\right|^2}{2(2\gamma+\Delta(H_A^{\prime}))^2} \leq \delta_{TC}^2 (\ket{\psi}) +  \delta_{CC}^2 (\ket{\psi}) \label{eq:uncertainty}.
\end{equation}

Next, we associate the mean squared on the right-hand side with the gate fidelity.  In the following calculation, we represent $\ket{\psi}$ as $\ket{\psi}_C =c_0 \ket{0}_C + c_1 \ket{1}_C$. Because  
\begin{eqnarray}
& & 1 = \Braket{\sigma(\vec{l^{\prime}})^{\prime 2}_C}= \Braket{\sigma(\vec{l})_C^2} =\sum_{a,b}  \left [|c_0|^2 |\ket{E_{ab}^{00}}|^2 +|c_1|^2 |\ket{E_{ab}^{10}}|^2  \right] ,
\\
& &\Braket{\sigma(\vec{l})_C\sigma(\vec{l^{\prime}})_C^{\prime}} = \sum_{a,b} (-1)^a  \left[c_0^* \Bra{E_{ab}^{00}} - c_1^* \Bra{E_{ab}^{10}} \right]\left[ c_0 \Ket{E_{ab}^{00} }+c_1 \Ket{E_{ab}^{10}} \right],  
	\end{eqnarray}
holds,  we obtain 
	\begin{eqnarray}
\delta_{CC}^2(\ket{\psi}) 
& =& 4\sin^2 \frac{\theta}{2} \left[ |c_0|^2 \left( \| \ket{E_{10}^{00}} \|^2+\| \ket{E_{11}^{00}} \|^2 \right) + |c_1|^2 \left( \| \ket{E_{00}^{10}}\|^2 + \| \ket{E_{01}^{10}}\|^2 \right)\right].
	\end{eqnarray}
Similarly, we associate  $\delta_{TC}^2 (\ket{\psi})$ with norms of external system states.  
	\begin{eqnarray}
& &\Braket{ \sigma(\vec{l})_C\sigma(\vec{v})_T^{\prime}} =  \sum_{a,b} (-1)^b  \left[c_0^* \Bra{E_{ab}^{00}} - c_1^*\Bra{E_{ab}^{10}} \right]\left[ c_0\Ket{E_{ab}^{00} }+c_1\Ket{E_{ab}^{10}} \right],  \\
& &\Braket{\sigma(\vec{l})_C\sigma(\vec{v} \times \vec{u})^{\prime}_T } 
= i \sum_{a,b} (-1)^{b+1}\left[c_0^* \Bra{E_{a,b}^{00}} - c_1^ * \Bra{E_{a,b}^{10}} \right]\left[ c_0 \Ket{E_{a,b+1}^{00} }+c_1 \Ket{E_{a,b+1}^{10}} \right],\\
& &\Braket{\sigma(\vec{v} \times \vec{u})_T^{\prime}\sigma(\vec{v})_T^{\prime} }  =  i \sum_{a,b} \left[c_0^* \Bra{E_{ab}^{00}} +c_1^* \Bra{E_{ab}^{10}} \right]\left[ c_0 \Ket{E_{a,b+1}^{00} }+c_1 \Ket{E_{a,b+1}^{10}} \right], 
\end{eqnarray}
are able to be shown. Therefore, 
	\begin{eqnarray}
& & \delta_{TC}^2 (\ket{\psi}) \nonumber \\
&=& \cos^2 \frac{\theta}{2} + 4\sin^2 \frac{\theta}{2} \left[|c_0|^2 \left(\|\ket{E_{01}^{00}}\|^2 +\|\ket{E_{11}^{00}}\|^2 \right)+|c_1|^2 \left( \|\ket{E_{00}^{10}} \|^2+\|\ket{E_{10}^{10}}\|^2 \right)\right]\nonumber \\
& & \quad -2i \sin \frac{\theta}{2} \cos \frac{\theta}{2} \sum_{a}   \left[ |c_0|^2 \Braket{E_{a0}^{00}|E_{a,1}^{00}}  - |c_1|^2 \Braket{E_{a,0}^{10}|E_{a,1}^{10}}\right]\nonumber \\
& &+2 i \sin \frac{\theta}{2} \cos \frac{\theta}{2} \sum_{a}  \left[ |c_0|^2\Braket{E_{a1}^{00}|E_{a,0}^{00}}  -|c_1|^2 \Braket{E_{a,1}^{10}|E_{a,0}^{10}}\right] \nonumber \\ \\
&\leq& 2|c_0|^2 \left[\cos \frac{\theta}{2} \Bra{E_{00}^{00}} +i\sin \frac{\theta}{2} \Bra{E_{01}^{00}} \right]\left[\cos \frac{\theta}{2} \ket{E_{00}^{00}} -i\sin \frac{\theta}{2} \ket{E_{01}^{00}} \right] \nonumber \\
& &+ 2 | c_0|^2 \left[\cos \frac{\theta}{2} \Bra{E_{10}^{00}} +i\sin \frac{\theta}{2} \Bra{E_{11}^{00}} \right]\left[\cos \frac{\theta}{2} \ket{E_{10}^{00}} -i\sin \frac{\theta}{2} \ket{E_{11}^{00}} \right] \nonumber \\
& & + 2 |c_1|^2\left[\cos \frac{\theta}{2} \Bra{E_{01}^{10}} +i\sin \frac{\theta}{2} \Bra{E_{00}^{10}} \right]\left[\cos \frac{\theta}{2} \ket{E_{01}^{10}} -i\sin \frac{\theta}{2} \ket{E_{00}^{10}} \right]\nonumber \\
& &+2|c_1|^2 \left[\cos \frac{\theta}{2} \Bra{E_{11}^{10}} +i\sin \frac{\theta}{2} \Bra{E_{10}^{10}} \right]\left[\cos \frac{\theta}{2} \ket{E_{11}^{10}} -i\sin \frac{\theta}{2} \ket{E_{10}^{10}} \right] \nonumber \\
& &+ 2 \left[|c_0|^2 \left( \|\Ket{E_{01}^{00}} \|^2 +\|\Ket{E_{11}^{00}}\|^2\right)+ |c_1|^2 \left( \|\Ket{E_{00}^{10}} \|^2+\|\Ket{E_{10}^{10}} \|^2 \right)\right] 
	\end{eqnarray}
holds.  We can derive the following equation: 
	\begin{eqnarray}
& &\delta_{CC}^2 (\ket{\psi})+\delta_{TC}^2 (\ket{\psi}) \nonumber \\
&\leq&4\sin^2 \frac{\theta}{2} \left[ |c_0|^2 \left( \| \ket{E_{10}^{00}} \|^2+\| \ket{E_{11}^{00}} \|^2 \right) + |c_1|^2 \left( \| \ket{E_{00}^{10}}\|^2 + \| \ket{E_{01}^{10}}\|^2 \right)\right] \nonumber \\
& & + 2|c_0|^2 \left[\cos \frac{\theta}{2} \Bra{E_{00}^{00}} +i\sin \frac{\theta}{2} \Bra{E_{01}^{00}} \right]\left[\cos \frac{\theta}{2} \ket{E_{00}^{00}} -i\sin \frac{\theta}{2} \ket{E_{01}^{00}} \right]\nonumber \\
& &+ 2 | c_0|^2 \left[\cos \frac{\theta}{2} \Bra{E_{10}^{00}} +i\sin \frac{\theta}{2} \Bra{E_{11}^{00}} \right]\left[\cos \frac{\theta}{2} \ket{E_{10}^{00}} -i\sin \frac{\theta}{2} \ket{E_{11}^{00}} \right] \nonumber \\
& & + 2 |c_1|^2\left[\cos \frac{\theta}{2} \Bra{E_{01}^{10}} +i\sin \frac{\theta}{2} \Bra{E_{00}^{10}} \right]\left[\cos \frac{\theta}{2} \ket{E_{01}^{10}} -i\sin \frac{\theta}{2} \ket{E_{00}^{10}} \right] \nonumber \\
& &+2|c_1|^2 \left[\cos \frac{\theta}{2} \Bra{E_{11}^{10}} +i\sin \frac{\theta}{2} \Bra{E_{10}^{10}} \right]\left[\cos \frac{\theta}{2} \ket{E_{11}^{10}} -i\sin \frac{\theta}{2} \ket{E_{10}^{10}} \right] \nonumber \\
& &+ 2 \left[|c_0|^2 \left( \|\Ket{E_{01}^{00}} \|^2 +\|\Ket{E_{11}^{00}}\|^2\right)+ |c_1|^2 \left( \|\Ket{E_{00}^{10}} \|^2+\|\Ket{E_{10}^{10}} \|^2 \right)\right] \\
& \leq & 4 \left[ |c_0|^2 \left( \| \ket{E_{10}^{00}} \|^2+\| \ket{E_{11}^{00}} \|^2 \right) + |c_1|^2 \left( \| \ket{E_{00}^{10}}\|^2 + \| \ket{E_{01}^{10}}\|^2 \right)\right] \nonumber \\
& & - 2|c_0|^2 \left[\sin \frac{\theta}{2} \Bra{E_{00}^{00}} -i\cos \frac{\theta}{2} \Bra{E_{01}^{00}} \right]\left[\sin \frac{\theta}{2} \ket{E_{00}^{00}} +i\cos \frac{\theta}{2} \ket{E_{01}^{00}} \right]\nonumber \\
& & + 2 |c_0|^2 \left(\|\ket{E_{00}^{00}} \|^2 + \|\ket{E_{01}^{00}} \|^2   \right)  + 2 | c_0|^2 \left[\cos \frac{\theta}{2} \Bra{E_{10}^{00}} +i\sin \frac{\theta}{2} \Bra{E_{11}^{00}} \right]\left[\cos \frac{\theta}{2} \ket{E_{10}^{00}} -i\sin \frac{\theta}{2} \ket{E_{11}^{00}} \right]  \nonumber \\
& &+ 2 |c_1|^2\left[\cos \frac{\theta}{2} \Bra{E_{01}^{10}} +i\sin \frac{\theta}{2} \Bra{E_{00}^{10}} \right]\left[\cos \frac{\theta}{2} \ket{E_{01}^{10}} -i\sin \frac{\theta}{2} \ket{E_{00}^{10}} \right] +2 |c_1|^2 \left( \|\ket{E_{11}^{10}} \|^2 +\|\ket{E_{10}^{10}} \|^2\right) \nonumber \\
& & -2|c_1|^2 \left[\sin \frac{\theta}{2} \Bra{E_{11}^{10}} -i\cos \frac{\theta}{2} \Bra{E_{10}^{10}} \right]\left[\sin \frac{\theta}{2} \ket{E_{11}^{10}} +i\cos \frac{\theta}{2} \ket{E_{10}^{10}} \right]  \nonumber \\
& &+ 2 \left[|c_0|^2 \left( \|\Ket{E_{01}^{00}} \|^2 +\|\Ket{E_{11}^{00}}\|^2\right)+ |c_1|^2 \left( \|\Ket{E_{00}^{10}} \|^2+\|\Ket{E_{10}^{10}} \|^2 \right)\right] \nonumber \\ \\
& =& 2|c_0|^2 - 2|c_0|^2 \left[\sin \frac{\theta}{2} \Bra{E_{00}^{00}} -i\cos \frac{\theta}{2} \Bra{E_{01}^{00}} \right]\left[\sin \frac{\theta}{2} \ket{E_{00}^{00}} +i\cos \frac{\theta}{2} \ket{E_{01}^{00}} \right] \nonumber \\
& & + 2 | c_0|^2 \left[\cos \frac{\theta}{2} \Bra{E_{10}^{00}} +i\sin \frac{\theta}{2} \Bra{E_{11}^{00}} \right]\left[\cos \frac{\theta}{2} \ket{E_{10}^{00}} -i\sin \frac{\theta}{2} \ket{E_{11}^{00}} \right] + 2|c_0|^2 \left(1 - \|\ket{E_{00}^{00}} \|^2 + \|\ket{E_{11}^{00}} \|^2\right)\nonumber \\ 
& & +2 |c_1|^2 + 2 |c_1|^2\left[\cos \frac{\theta}{2} \Bra{E_{01}^{10}} +i\sin \frac{\theta}{2} \Bra{E_{00}^{10}} \right]\left[\cos \frac{\theta}{2} \ket{E_{01}^{10}} -i\sin \frac{\theta}{2} \ket{E_{00}^{10}} \right] \nonumber \\ 
& & -2|c_1|^2 \left[\sin \frac{\theta}{2} \Bra{E_{11}^{10}} -i\cos \frac{\theta}{2} \Bra{E_{10}^{10}} \right]\left[\sin \frac{\theta}{2} \ket{E_{11}^{10}} +i\cos \frac{\theta}{2} \ket{E_{10}^{10}} \right] +2 |c_1|^2 \left( 1 - \|\ket{E_{11}^{10}} \|^2 + \|\ket{E_{00}^{10}} \|^2\right) \nonumber \\ \\
&\leq& 4|c_0|^2[1- f(\ket{0,0})^2]+4|c_1|^2[1-F(\ket{1,0})^2]+4|c_0|^2[1-F(\ket{0,0})^2]+4|c_1|^2[1-f(\ket{1,0})^2] \\
& \leq & 4[1-F(\mathcal{E}_{\alpha},U_{ideal})^2]+4[1-f(\mathcal{E}_{\alpha},U_{ideal})^2].
	\end{eqnarray} 

Imposing the symmetry that the gate fidelity when the target gate is $V_T(\theta)$ is equal to the gate fidelity when the  target gate is $V_T(-\theta)$ , $F(\mathcal{E}_{\alpha},U_{ideal})=f(\mathcal{E}_{\alpha},U_{ideal})$, 
\begin{equation}
\sin^2 \frac{\theta}{2}\cdot \frac{\left|\Braket{\left[\sigma(\vec{l})_C,H_C\right]}\right|^2}{2(2\gamma+\Delta(H^{\prime}_A))^2} \leq   8[1-F(\mathcal{E}_{\alpha},U_{ideal})^2] 
\end{equation}
can be proved.  Finally, by maximizing over states of $C$+$T$ and minimizing over states of $A$,   Eq.~(\ref{eq:1bit}) is derived.
\end{proof}

In this section, we show that $H_C$ must satisfy Eq.~(\ref{eq:wayone_constraint}) and Eq.~(\ref{eq:waytwo_contraint}) to implement a non-trivial controlled gate whose action is given by Eq.(\ref{eq:controlled-U}) with zero-error  using scattering process under energy conservation law. We also obtain the upper bound of gate fidelity Eq.~(\ref{eq:1bit}) for a two-qubit controlled gate.


\section{Summary}
We showed a lower bound of  the measurement error Eq.~(\ref{eq:way_energy}) for scattering-type measurements   which satisfy Eq.~(\ref{eq:assump1})  under the energy conservation in section 2.  In section 2 we also presented that the SWAP gate is not able to be implemented perfectly as an application of Eq.~(\ref{eq:way_energy}).  Futhermore,  the bound of the gate fidelity of the Hadamard gate proved in Ref. \cite{Ozawa-way} under an additive conservation law is able to be expanded under the energy conservation in scattering processes. In section 3,  we gived the necessary condition Eq.  (\ref{eq:wayone_constraint}) which must be fulfilled to implement a controlled unitary gate described by Eq.  (\ref{eq:controlled-U})  without error in a scattering process under energy conservation. We also proved that when the interaction term satisfies Eq.  (\ref{eq:assump2}), the additional necessary condition for $H_C$, Eq.~(\ref{eq:waytwo_contraint}) must also be fulfilled to implement a controlled unitary gate perfectly. Moreover,  when $C$ and $T$ are one-qubit systems, the quantitative  relation Eq.  (\ref{eq:1bit}) between the gate fidelity and the non-diagonal entry of $H_C$ is shown. This result extends Ozawa's result which evaluates the upper bound of the gate fidelity of CNOT gates \cite{Ozawa-CNOT}.


\begin{acknowledgments}
This research was partially supported
by JSPS KAKENHI Grants 
No.~JP24H01566 (M.O.),
No.~JP22K03424 (M.O.), 
No.~JP21K11764 (M.O.), 
No.~JP19H04066 (M.O.),
No.~JP19K03838 (M.H.),
No.~JP21H0518 (M.H.),
No.~JP21H05188 (M.H.),
JST CREST Grant Number JPMJCR23P4 (M.O.), Japan,
Foundational Questions Institute (M.H.), Silicon Valley
Community Foundation (M.H.), JST SPRING, Grant
No.~JPMJSP2114 (R.K.), a Scholarship of Tohoku
University, Division for Interdisciplinary Advanced
Research and Education (R.K.), and the WISE Program
for AI Electronics, Tohoku University (R.K.).
\end{acknowledgments}

\bibliography{reference}
\end{document}